\documentclass{tlp}

\title{A Proof Theoretic Approach to Failure in Functional Logic Programming
}

\usepackage{latexsym}
\usepackage{amsfonts}
\usepackage{amssymb}

\newtheorem{definition}{Definition} % [section]
\newtheorem{example}{Example} % [section]
\newtheorem{lemma}{Lemma} % [section]
\newtheorem{proposition}{Proposition} % [section]
\newtheorem{theorem}{Theorem} % [section]
\newtheorem{corollary}{Corollary} % [section]

\newcommand{\coment}[1]{}

\newcommand{\eqr}{\rightarrow}  % igualdad de las reglas
\newcommand{\nat}{\mathrm{I\! N}}
\newcommand{\var}{\mathcal{V}}

\newcommand{\con}{\bowtie}
\renewcommand{\div}{\mathrel{\mathord{<}\mkern-4mu\mathord{>}}}
\newcommand{\nocon}{\not\bowtie}
\newcommand{\nodiv}{\mathrel{\mathord{<}\mkern-6mu\mathord{/}\mkern-6mu\mathord{>}}}

\newcommand{\rec}{\vartriangleleft}
\newcommand{\recr}{\vartriangleleft_{_R}}
\newcommand{\fail}{\textsf{\scriptsize F}}

\newcommand{\toye}{$\mathcal{TOY\;\,}$}
\newcommand{\toy}{$\mathcal{TOY}$}

\newcommand{\cb}{\mathcal{C}}

\newcommand{\conscrwl}{\vdash_{\textit{\tiny CRWL}}}
\newcommand{\conscrwlf}{\vdash_{\textit{\tiny CRWLF}}}

\newcommand{\mifrac}[2]{\frac{\mbox{$
          \begin{array}{c}
            \\[-0.4cm]
            #1\\
            \\[-0.4cm]
          \end{array}$}}
      {\mbox{$
          \begin{array}{c}
            \\[-0.4cm]
            #2\\
            \\[-0.4cm]
          \end{array}$}}}

\newcommand{\ca}{\mbox{\tiny{3}}\! \frac{}{\mbox{$
      \begin{array}{c}
        \\[-0.4cm]
        z\to z\\
        \\[-0.4cm]
      \end{array}$}}}
\newcommand{\cbb}{\mbox{\tiny{4}}\! \frac{\ca}{\mbox{$
      \begin{array}{c}
        \\[-0.5cm]
        coin\to z
        \\[-0.3cm]
      \end{array}
$}}}
\newcommand{\cc}{\mbox{\tiny{4}}\! \frac{\ca\quad \ca\quad \ca}{\mbox{$add(z,z)\to z$}}}
\newcommand{\cd}{\mbox{\tiny{4}}\! \frac{\cbb\quad \cc}{\mbox{$double(coin)\to z$}}}

\newcommand{\re}{\rec}

\newcommand{\fa}{\mbox{\tiny{3}}\! \frac{}{\mbox{$
      \begin{array}{c}
        \\[-0.3cm]
        z\re\{z\}\\
        \\[-0.3cm]
      \end{array}
$}}}

\newcommand{\fnbaaa}{\mbox{\tiny{1}}\! \frac{}{\mbox{$
      \begin{array}{c}
        \\[-0.3cm]
        z\re\{ \bot\}\\
        \\[-0.3cm]
      \end{array}
$}}}

\newcommand{\fnbaa}{\mbox{\tiny{3}}\! \frac{\fnbaaa}{\mbox{$
      \begin{array}{c}
        \\[-0.3cm]
        s(z)\re\{ s(\bot)\}\\
        \\[-0.3cm]
      \end{array}
$}}}

\newcommand{\fnba}{\mbox{\tiny{6}}\! \frac{\fnbaa}{\mbox{$
      \begin{array}{c}
        \\[-0.3cm]
        h\re_{_{1\!}}\{ s(\bot)\}\\
        \\[-0.3cm]
      \end{array}
$}}}

\newcommand{\fnbbbb}{\mbox{\tiny{1}}\! \frac{}{\mbox{$
      \begin{array}{c}
        \\[-0.3cm]
        h\re\{ \bot\}\\
        \\[-0.3cm]
      \end{array}
$}}}

\newcommand{\fnbbb}{\mbox{\tiny{3}}\! \frac{\fnbbbb}{\mbox{$
      \begin{array}{c}
        \\[-0.3cm]
        s(h)\re\{ s(\bot)\}\\
        \\[-0.3cm]
      \end{array}
$}}}

\newcommand{\fnbb}{\mbox{\tiny{6}}\! \frac{\fnbbb}{\mbox{$
      \begin{array}{c}
        \\[-0.3cm]
        h\re_{_{2\!}}\{ s(\bot)\}\\
        \\[-0.3cm]
      \end{array}
$}}}

\newcommand{\fnb}{\mbox{\tiny{4}}\! \frac{\fnba\quad \fnbb}{\mbox{$
      \begin{array}{c}
        \\[-0.3cm]
        h\re\{ s(\bot)\}\\
        \\[-0.3cm]
      \end{array}
$}}}

\newcommand{\fna}{\mbox{\tiny{3}}\! \frac{\fnb}{\mbox{$
      \begin{array}{c}
        \\[-0.3cm]
        s(h)\re\{ s(s(\bot))\}\\
        \\[-0.3cm]
      \end{array}
$}}}

\newcommand{\fnxt}{\mbox{\tiny{3}}\! \frac{}{\mbox{$
      \begin{array}{c}
        \\[-0.3cm]
        [\ ]\re\{ [\ ]\}\\
        \\[-0.3cm]
      \end{array}
$}}}

\newcommand{\fn}{\mbox{\tiny{3}}\! \frac{\fna \quad \fnxt}{\mbox{$
      \begin{array}{c}
        \\[-0.3cm]
        [s(h)]\re\{ [s(s(\bot))]\}\\
        \\[-0.3cm]
      \end{array}
$}}}

\newcommand{\fo}{\mbox{\tiny{3}}\! \frac{}{\mbox{$
      \begin{array}{c}
        \\[-0.3cm]
        z\re\{z\}\\
        \\[-0.3cm]
      \end{array}
$}}}
\newcommand{\fp}{\mbox{\tiny{1}}\! \frac{}{\mbox{$
      \begin{array}{c}
        \\[-0.3cm]
        \bot\re\{\bot\}\\
        \\[-0.3cm]
      \end{array}
$}}}
\newcommand{\fq}{\mbox{\tiny{3}}\! \frac{\fp}{\mbox{$
      \begin{array}{c}
        \\[-0.3cm]
        s(\bot)\re\{s(\bot)\}\\
        \\[-0.3cm]
      \end{array}
      $}}}
\newcommand{\fr}{\mbox{\tiny{3}}\!
  \frac{\fq}{\mbox{$
      \begin{array}{c}
        \\[-0.3cm]
        s(s(\bot))\re\{s(s(\bot))\}\\
        \\[-0.3cm]
      \end{array}
      $}}}
\newcommand{\fs}{\mbox{\tiny{11}}\! \frac{\fo\quad \fr}{\mbox{$
      \begin{array}{c}
        \\[-0.3cm]
        z\nocon s(s(\bot))\\
        \\[-0.3cm]
      \end{array}
$}}}
\newcommand{\ft}{\mbox{\tiny{7}}\! \frac{\fs}{\mbox{$
      \begin{array}{c}
        \\[-0.3cm]
        \varphi_3\equiv mb(z,[s(s(\bot))])\re_{_1\!}\{\fail\}\\
        \\[-0.3cm]
      \end{array}
$}}}

\newcommand{\fu}{\mbox{\tiny{3}}\! \frac{}{\mbox{$
      \begin{array}{c}
        \\[-0.3cm]
        [\ ]\re\{ [\ ] \}\\
        \\[-0.3cm]
      \end{array}
      $}}}
\newcommand{\fv}{\mbox{\tiny{8}}\! \frac{}{\mbox{$
      \begin{array}{c}
        \\[-0.3cm]
        mb(z,[\ ])\re_{_{1\!}}\{\fail\}\\
        \\[-0.3cm]
      \end{array}
      $}}}

\newcommand{\fvv}{\mbox{\tiny{8}}\! \frac{}{\mbox{$
      \begin{array}{c}
        \\[-0.3cm]
        mb(z,[\ ])\re_{_{2\!}}\{\fail\}\\
        \\[-0.3cm]
      \end{array}
      $}}}

\newcommand{\fx}{\mbox{\tiny{4}}\! \frac{\fa\ \fu\ \fv\ \fvv}{\mbox{$
      \begin{array}{c}
        \\[-0.3cm]
        mb(z,[\ ])\re\{\fail\}\\
        \\[-0.3cm]
      \end{array}
      $}}}

\newcommand{\fz}{\mbox{\tiny{11}}\! \frac{\fx\ \fy}{\mbox{$
      \begin{array}{c}
        \\[-0.3cm]
        mb(z,[\ ])\nocon\{\mathsf{t}\}\\
        \\[-0.3cm]
      \end{array}
      $}}}

\newcommand{\fy}{\mbox{\tiny{3}}\! \frac{}{\mbox{$
      \begin{array}{c}
        \\[-0.3cm]
        \mathsf{t}\re\{ \mathsf{t}\}\\
        \\[-0.3cm]
      \end{array}
      $}}}

\newcommand{\faa}{\mbox{\tiny{7}}\!\frac{\fz}{\mbox{$
      \begin{array}{c}
        \\[-0.3cm]
        \varphi_4\equiv mb(z,[s(s(\bot))])\re_{_2\!}\{\fail\}\\
        \\[-0.3cm]
      \end{array}
      $}}}

\author[F.~J. L\'opez-Fraguas and J. S\'anchez-Hern\'andez]{FRANCISCO J.  L\'OPEZ-FRAGUAS and JAIME S\'ANCHEZ-HERN\'ANDEZ
              \thanks{The authors  have been partially supported by the Spanish CICYT (project TIC 2002-01167
                                 `MEHLODIAS').}
 \\Dep. Sistemas Inform\'{a}ticos y Programaci\'{o}n,
Univ. Complutense de Madrid\\ E-mail: $\{$fraguas,jaime$\}$@sip.ucm.es}

\begin{document}

\maketitle

\begin{abstract}
How to extract negative information from programs is an important issue in 
logic programming. Here  we address the problem for functional logic
programs, from a proof-theoretic perspective. 
The starting point of our work is {\it CRWL} (Constructor based ReWriting Logic), a
well established theoretical framework  for functional logic programming, 
whose fundamental notion
is that of  non-strict non-deterministic function. We present a proof
calculus, {\it CRWLF},  which is able to deduce negative information from {\it CRWL}-programs.
In particular, {\it CRWLF} is able to prove `finite' failure of reduction within {\it CRWL}.
\end{abstract}

Submitted: Dec 22, 2000, revised: Feb 21, 2002, accepted: Sept 27 2002.

\section{Introduction}
We address in this paper the problem of extracting negative information 
from functional logic programs.
The question of negation is a main topic of research in the logic programming field,
and the most common approach is {\em negation as failure} \cite{clark78},
as an easy effective approximation to the {\it CWA} {(\em closed world assumption}), which
is a simple, but uncomputable,  way of deducing negative information from positive
programs (see e.g. \citeN{AptBol94} for a survey on negation in logic programming).

On the other hand, functional logic programming ({\it FLP} for short) is a powerful programming
paradigm trying to combine the nicest properties of functional and logic
programming (see \citeN{Hanus94JLP} for a now `classical' survey on {\it FLP}).
A mainstream in current {\it FLP}  research considers languages
which are biased to the functional programming  style, in the sense that programs  define
functions, but having logic programming capabilities because their operational mechanisms
are based on narrowing. Some existing systems of this kind are \toye
\cite{LS99a,Toy} or the various implementations of {\it Curry}  \cite{Han99a}.
In the rest of the paper we have in mind such approach when we
refer to {\it FLP}.

{\it FLP} subsumes {\em pure} logic programming: predicates
can be defined as functions returning the value `true', for which definite
clauses can be written as conditional rewrite rules. In some simple cases
it is enough, to handle negation, just to define predicates as two-valued boolean functions
returning the values `true' or `false'.
But negation as failure is far more expressive, as we see in the next section,
and it is then of clear interest to investigate a similar notion  for the case
of {\it FLP}. Failure in logic programs, when seen as functional logic
programs, corresponds to failure of reduction to `true'. This generalizes
to a natural notion  of failure in {\it FLP},  which is 
`failure of reduction to (partial) data constructor value', or in other terms,
`failure of reduction to head normal form' ({\it hnf} for short).

As technical setting for our work we have chosen {\it CRWL} \cite{GHL96,GHL99}, a 
well established theoretical framework  for {\it FLP}. 
The fundamental notion in {\it CRWL}
is that of  non-strict non-deterministic function, for which 
{\it CRWL} provides a firm logical basis. Instead of
equational logic, which is argued to be unsuitable for {\it FLP} in
\citeN{GHL99}, {\it CRWL} considers a Constructor based ReWriting Logic, presented by means of a
proof calculus, which determines what statements can be deduced from a given program.
In addition to the proof-theoretic semantics, \cite{GHL96,GHL99} develop a model theoretic
semantics for  {\it CRWL}, with existence of distinguished free term models for programs, and a
sound  and complete lazy narrowing calculus as operational semantics. 
The interest of {\it CRWL} as a theoretical framework for {\it FLP} has been
mentioned in \citeN{Han99a}, and is further evidenced by its many extensions
incorporating relevant aspects of declarative programming like HO
features \cite{GHR97}, polymorphic and algebraic types \cite{AR01},
or constraints \cite{ALR98}. The framework, with many of these extensions
(like types, HO and constraints) has been implemented in the system \toy. 

Here we are interested in extending the proof-theoretic side of {\it CRWL} to
cope with failure. More concretely, we look for a proof calculus, which will
be called {\it CRWLF} (`{\it CRWL} with failure'), which is able to prove failure of 
reduction in
{\it CRWL}. Since reduction in {\it CRWL} is expressed by proving certain statements,
our calculus will provide proofs of unprovability within {\it CRWL}. As for the case
of {\em CWA}, unprovability is not computable, which means that our calculus can only
give an approximation, corresponding to cases which can be intuitively described
as `finite failures'.

There are very few works about negation in {\it FLP}. In \citeN{Moreno-Navarro94} the work
of Stuckey about {\em constructive negation} \cite{Stuckey91,Stuckey95} is adapted to the
case of {\it FLP} with strict functions and innermost narrowing as operational mechanism.
In \citeN{Moreno96ELP} a similar work is done for the case of non-strict functions
and lazy narrowing. The approach is very different of the proof-theoretic view
of our work. The fact that we also consider non-deterministic functions makes a significant
difference.

The proof-theoretic approach, although not very common, has been followed sometimes
in the logic programming field, as in \citeN{jager98}, which develops for logic
programs (with negation) a framework which resembles, in a very general sense, 
${\it CRWL}$: a program determines a deductive system for which deducibility,  
validity in a class of models, validity in a distinguished model and derivability
by an operational calculus are all equivalent. Our work attempts to be the first
step of what could be a similar programme for {\it FLP} extended with the use of
failure when writing programs.

The rest of the paper is organized as follows. In Section 2 we discuss the 
interest of using failure as a programming construct in the context of {\it FLP} .
In Section 3 we give the
essentials of {\it CRWL} which are needed for our work. Section 3 presents 
the {\it CRWLF}-calculus, preceded by some illustrative examples.
Sections 4, 5 and 6 constitute the technical core of the paper,
presenting the properties of {\it CRWLF}
and its relation to {\it CRWL}.
Finally, Section 7 outlines some conclusions and possible future work.

\section{The Interest of Failure in {\it FLP}}
\label{sec:interest}

Although this work is devoted only to the theoretical aspects of failure in {\it FLP}, in this
section we argue some possible applications of this resource from the
point of view of writing functional logic programs.

{\it FLP} combines some of the main capabilities of the two
main streams of declarative programming: functional programming ({\it FP}) and
logic programming ({\it LP}). Theoretical aspects of {\it FLP} are well established
(see e.g. \citeN{GHL99}) and there are also practical implementations such as {\it Curry} or \toy.  Disregarding syntax, both pure Prolog and (a wide subset of) Haskell
are subsumed by those systems. The usual claim is then that by the use of an {\it FLP}  system
one can choose the style of programming better suited to each occasion.

However there are features related to failure, mainly in {\it
  LP} (but also in {\it FP}) yet not
available in {\it FLP} systems.  This poses some problems to {\it FLP}: if
a logic  program uses negation (a very common situation), it cannot be seen as
an {\it FLP}  program. This is not a very serious inconvenience if other features
of {\it FLP}  could easily replace the use of failure. But
if the {\it FLP}  solution  (without failure) to a problem is significantly more complex than, say,  an {\it LP}   solution
making use of failure,  then it is not worth to use {\it FLP}  for that
problem, thus contradicting in practice the claim that  {\it FLP}   can successfully  replace {\it LP}  and {\it FP} .

We now give concrete examples of the potential use of a construction to express
failure in {\it FLP}  programs.
We assume for the examples below that we incorporate to {\it FLP} the following
function  to express failure of an expression:

\begin{center}
  ${\it fails}(e)::=
  \left\{ \begin{array}{ll}
    {\it true} & \textrm{if $e$ fails to be reduced to hnf}\\
    {\it false} & \textrm{otherwise}
  \end{array}\right.$
\end{center}
The sensible notion to consider is {\em failure of reduction to head normal form}
\footnote{To be technically more precise, we should speak of `failure to
reduction to head normal form with respect to the {\em CRWL}-calculus', to be recalled in
Section 3.},
since head normal forms
(i.e., variables or expressions $c(\ldots)$, where $c$ is a constructor symbol) are the expressions
representing, without the need of further reduction, defined (maybe partial) values.

\begin{example}
 [Failure to express negation in  {\it LP}]
\vspace*{0.2cm}

\noindent
The most widespread approach to negation in the {\it LP} paradigm is {\it negation
  as failure} \cite{clark78}, of which 
all {\it PROLOG} systems provide an
implementation. Typically, in a logic program one writes clauses defining
 the positive cases for a predicate, and the effect of using negation is to `complete'
the definition with the negative cases, which correspond to failure of the given clauses.

For example, in {\it LP} the
predicate {\it member} can be defined
as:
\begin{center}
  $\begin{array}{l}
    member(X,[X|Ys]).\\
    member(X,[Y|Ys])\leftarrow member(X,Ys).
  \end{array}$
\end{center}
This defines {\it member(X,L)} as a semidecision procedure to check if $X$ is an element of
$L$. If one needs to check that $X$ is not an element of $L$, then negation can be used, as in
the clause
\begin{center}
  $add(X,L,[X|L]) :- not\  member(X,L).$
\end{center}

Predicates like {\it member} can be defined in {\it FLP}  as {\it true}-valued functions,
converting clauses into conditional rules returning {\it true}:
\begin{center}
  $\begin{array}{l}
    member(X,[Y|Ys]) \eqr true \Leftarrow X \con Y\\
    member(X,[Y|Ys]) \eqr true \Leftarrow member(X,Ys)\con true
  \end{array}$
\end{center}

To achieve  linearity (i.e., no variable repetition) of heads, a usual
 requirement in {\it FLP}, the condition $X \con Y$ is used in the first rule.
 The symbol $\con$ (taken from \cite{GHL96,GHL99}) is used throughout the
paper to express `joinability', which means that both sides can be reduced to the
same data value (for the purpose of this example, $\con$  can be read simply as
strict equality).

What cannot be directly translated into {\it FLP}  (without failure) is a clause like that
of {\it add}, but with failure it is immediate:
\begin{center}
  $add(X,L,[X'|L']) \eqr true \Leftarrow  {\it fails}(member(X,L)) \con true, X' \con X, L' \con L$
\end{center}

In general, any literal of the form {\it not Goal} in a logic program can be
replaced  by ${\it fails}(Goal) \con true$ in its {\it FLP}-translation.

This serves to argue that {\it FLP}  with failure subsumes {\it LP}  with negation, but of
course this concrete example corresponds to the category of `dispensable' uses
of failure, because there is
a natural failure-free {\it FLP}  counterpart to the predicate {\it member}
in the form of a bivaluated boolean function, where the
failure is expressed by the value {\it false}. The following  could be such a  definition of
{\it member}:
\begin{center}
  $\begin{array}{lcl}
    member(X,[\ ]) & \rightarrow & {\it false}\\
    member(X,[Y|Ys]) & \rightarrow & true \Leftarrow X \con Y\\
    member(X,[Y|Ys]) & \rightarrow & member(X,Ys)\Leftarrow X \div Y
  \end{array}$
\end{center}

The symbol $\div$ (corresponding to disequality $\not=$ of \cite{LS99a,LS99b})
expresses `divergence', meaning that both sides can be reduced to some extent as to
detect inconsistency, i.e., conflict of constructors at the same position (outside function
applications). Now {\it add} can be easily defined without using failure:
\begin{center}
  $add(X,L,[X'|L']) \eqr true \Leftarrow member(X,L) \con false, X' \con X, L' \con L$
\end{center}

The next examples show situations where the use of negation is more `essential',
in the sense that it is  the natural way (at least a very natural way) of doing things.
\end{example}

\begin{example}[Failure in search problems I]
\label{grafo}
\vspace*{0.2cm}

\noindent
Non-deterministic constructs are a useful way of
programming problems involving search. In {\it FLP} one can choose to use predicates,
as in {\it LP}, or non-deterministic functions.
 In these cases, the use of failure can greatly simplify the task of programming.
We see an example with non-deterministic functions, a quite specific {\it FLP}  feature
which is known to be useful for programming \cite{Toy,Han99a,Ant97} in systems like {\it Curry} or \toy.

Consider  the problem of deciding, for acyclic directed graphs,
if there is a path connecting two nodes. A graph can be represented by a
non-deterministic function {\it next}, with rules of 
the form $next(N) \rightarrow N'$, indicating that there is an arc from $N$ to $N'$.
A concrete graph with nodes {\it a}, {\it b}, {\it c} and {\it d} could be given by the rules:

\begin{center}
  $next(a)\rightarrow b$\\
  $next(a)\rightarrow c$\\
  $next(b)\rightarrow c$\\
  $next(b)\rightarrow d$\\
\end{center}

and to determine if there is a path from $X$ to $Y$ we can define:

\begin{center}
  $\begin{array}{l}
    path(X,Y)\rightarrow true \Leftarrow X \con Y\\
    path(X,Y)\rightarrow true \Leftarrow X \div Y ,\ path(next(X),Y) \con true
  \end{array}$
\end{center}

Notice that {\it path} behaves as a semidecision
procedure recognizing only the positive cases, and there is no clear way (in
`classical' {\it FLP} ) of completing its definition with the negatives ones, unless
we change from the scratch  the representation of graphs. Therefore we cannot, for instance,
program in a direct way a property like
\begin{center}
{\it safe(X) ::= X is not connected with d}
\end{center}

Using failure this is an easy task:
\begin{center}
 $ {\it safe}(X) \rightarrow {\it fails}(path(X,d))$
\end{center}

With this definition, ${\it safe}(c)$ becomes $true$, while
${\it safe}(a)$, ${\it safe}(b)$ and ${\it safe}(d)$ are all $false$.
\end{example}

\begin{example} [Failure in search problems II]
\label{nim}
\vspace*{0.2cm}

\noindent
We examine now an example mentioned  in \citeN{Apt00} as one
striking illustration of the power of failure as expressive
resource in {\it LP} .
We want to program a two-person finite game where the players must perform
alternate legal moves, until one of them, the loser, cannot move.

We assume that legal moves from a given state
are programmed by a non-deterministic function
{\it move(State)} returning the new state after the movement.
Using failure it is easy to program a function to perform a winning
movement from a given position, if there is one:
\begin{center}
  $\begin{array}{ll}
    winMove(State) \eqr State' \Leftarrow & State' \con move(State),\\
    & {\it fails}(winMove(State')) \con true
  \end{array}$
\end{center}
We think it would be difficult to find a simpler coding without using failure.

As a concrete example we consider the well-known game {\it Nim}, where there are
some rows of sticks, and each player in his turn
must pick up one or more sticks from one of the rows.
A player loses when he cannot make a movement, that is,
when there are not more sticks  because the other player (the winner) has picked up the
last one.
 Nim states can be defined
by a list of natural numbers (represented by $0$ and $s(\_)$ as usual), and
the non-deterministic function $move$ can be programmed as:

\begin{center}
 $\begin{array}{l}
 move([N| Ns]) \eqr [pick(N) | Ns]\\
 move([N|Ns]) \eqr [N | move(Ns)]\\[1ex]
 pick (s(N)) \eqr N\\
 pick (s(N)) \eqr pick(N)
\end{array}$
\end{center}

A winning move from the state $[s(s(z)),s(z)]$ can be obtained by reducing the
expression ${\it winMove}([s(s(z)),s(z)])$. The proof calculus presented in
Sect. \ref{sec:cal_crwl} can prove that it can be reduced to $[s(z),s(z)]$, and
it is easy to check that this move guarantees the victory.

\end{example}

\begin{example}[Failure to express default rules]
%{\bf Failure in {\it FP}}
\vspace*{0.2cm}

\noindent
Compared to the case of {\it LP}, failure is not a so important programming
construct in {\it FP}.
There is still one practical feature of existing {\it FP} languages somehow related to failure, which is
the possibility of defining functions with {\em default rules}.
 In many {\it FP} systems pattern matching
determines {\it the} applicable rule for a function call, and as rules are tried
from top to bottom, default rules are implicit in the definitions. In
fact, the $n+1$-th rule in a definition is only applied if the first $n$ rules
are not applicable. For example, assume the following definition for the
function $f$:
\begin{center}
  $\begin{array}{ll}
    f(0) & \to 0\\
    f(X) & \to 1\\
  \end{array}$
\end{center}
The evaluation of the expression $f(0)$ in a functional language like Haskell \cite{PH99},
will produce the value $0$ by the first rule. The second rule is not used
for evaluating $f(0)$, even if pattern matching would succeed if the rule would
be considered in isolation. This sequential treatment of rules is useful in some
cases, specially for writing `last' rules covering default cases whose direct formulation
with pattern matching could be complicated. But observe that in systems allowing such
sequential trials of pattern matching, rules  have not a
declarative meaning by themselves;
their interpretation depends also on the previous rules.

This contrasts with  functional logic
languages  which try to
preserve the declarative reading of each rule. In such systems the expression $f(0)$
of the example above is
reducible, by applying in a non-deterministic way any of the rules, to the values
$0$ and $1$.

To achieve (and generalize) the effect of default rules in {\it FLP}, an explicit syntactical
construction {\it 'default'} can be introduced, as it has been done in
\cite{Moreno-Navarro94}. The function $f$ could be defined as:
\begin{center}
$\begin{array}{l}
    f(0)\to 0\\
    default\ f(X) \to 1\\
  \end{array}$
\end{center}

The intuitive operational meaning is: to reduce a call to $f$ proceed with the
first rule for $f$; if the reduction fails then try the default rule.

The problem now is how to achieve this behavior while preserving
the equational reading of each rule.
Using conditional rewrite rules and our function  ${\it fails}(\_)$,
we can transform the definition of a function to eliminate default rules. In
the general case we can consider conditional rewrite rules for the original
definition. Let $h$ be a function defined as:

\begin{center}
  $
  \begin{array}{l}
    h(\overline{t}_1) \to e_1 \Leftarrow \overline{C}_1\\
    ...\\
    h(\overline{t}_n) \to e_n \Leftarrow \overline{C}_n\\
    default\ h(\overline{t}_{n+1}) \to e_{n+1} \Leftarrow \overline{C}_{n+1}\\
  \end{array}$
\end{center}

The idea of the transformation is to consider a new function $h'$ defined by the
first $n$ rules of $h$. The original $h$ will be defined as $h'$ if it succeeds
and as the default rule if $h'$ fails:

\begin{center}
  $
  \begin{array}{l}
    h(\overline{X}) \to h'(\overline{X})\\
    h(\overline{X}) \to e_{n+1} \Leftarrow {\it fails}(h'(\overline{X}))\con true,\ \overline{C}_{n+1}\\[0.2cm]
    h'(\overline{t}_1) \to e_1 \Leftarrow \overline{C}_1\\
    ...\\
    h'(\overline{t}_n) \to e_n \Leftarrow \overline{C}_n\\
  \end{array}$
\end{center}

Applying this transformation to our function example $f$, we obtain:

\begin{center}
$\begin{array}{l}
    f(X)\to f'(X)\\
    f(X)\to 1 \Leftarrow {\it fails}(f'(X)) \con true\\[0.2cm]
    f'(0) \to 0\\[0.2cm]
 \end{array}$
\end{center}

With this definition we have got the expected behavior for $f$ without losing
the declarative reading of rules.

As another example, we can use a default rule to complete  the definition of the
function {\it path} in the example \ref{grafo} above:

\begin{center}
  $\begin{array}{l}
    path(X,Y)\eqr true \Leftarrow X \con Y\\
    path(X,Y)\eqr true \Leftarrow X \div Y , path(next(X),Y) \con true\\
    default\ path(X,Y)\eqr {\it false}
  \end{array}$
\end{center}

The function {\it safe} can now be written as:
\[
  {\it safe}(X) \rightarrow neg(path(X,d))
\]
where $neg$ is the boolean function 
\begin{center}
  $\begin{array}{ll}
    neg(true) & \eqr {\it false}\\
    neg(false) & \eqr {\it true}
  \end{array}$
\end{center}

Notice that in this example the (implicit) condition for applying the default rule
of {\it path} is far more complex than a merely syntactical default case expressing failure of
pattern matching, a feature recently discussed in \cite{currylist} as useful for {\it
  FLP}. Of course, default rules in the sense of \cite{Moreno-Navarro94} and of
this paper also cover such syntactical cases.

\end{example}

\section{The {\it CRWL} Framework}

We give here a short summary of
(a slight variant of) {\it CRWL}, in its proof-theoretic face.
Model theoretic semantics and lazy narrowing operational semantics
are not considered here. Full details can be found in \cite{GHL99,LS99b}.

\subsection{Technical Preliminaries}
\label{sec:crwl_prelim}

We assume a signature $\Sigma =DC_{\Sigma}\cup FS_{\Sigma}$ where
$DC_{\Sigma}=\bigcup_{n\in \nat} DC_{\Sigma}^n$ is a set of {\em
constructor} symbols and $FS_{\Sigma}=\bigcup_{n\in\nat}
FS_{\Sigma}^n$ is a set of \emph{function} symbols, all of them with
associated arity and such that $DC_{\Sigma}\cap
FS_{\Sigma}=\emptyset$. We also assume a countable set $\var$ of
\emph{variable} symbols. We write $Term_{\Sigma}$ for the set of
(total) {\em terms} (we say also {\em expressions}) built up with $\Sigma$ and $\var$ in the usual
way, and we distinguish the subset $CTerm_{\Sigma}$ of (total) constructor
terms or (total) \emph{c-terms}, which only make use of $DC_{\Sigma}$
and $\var$. The subindex $\Sigma$ will usually be omitted.
Terms intend to represent possibly reducible expressions,
while c-terms represent data values, not further reducible.

We will need sometimes to use the signature $\Sigma_{\bot}$ which is the result
of extending $\Sigma$ with the new constant (0-arity constructor) $\bot$, that
plays the role of the undefined value. Over $\Sigma_{\bot}$, we can build up the
sets $Term_\bot$ and $CTerm_\bot$ of (partial) terms and (partial) c-terms
respectively. Partial c-terms represent the result of partially evaluated
expressions; thus, they can be seen as approximations to the value
of expressions.

As usual notations we will write $X,Y,Z,...$ for variables, $c,d$ for
constructor symbols, $f,g$ for functions, $e$ for terms and $s,t$
for c-terms. In all cases, primes (') and subindices can be used.

We will use the sets of substitutions $CSubst=\{ \theta:\var\to
CTerm\}$ and $CSubst_{\bot}=\{ \theta:\var\to CTerm_\bot\}$.  We write $e\theta$
for the result of applying $\theta$ to  $e$.

Given a set of constructor symbols $S$ we say that the
c-terms
$t$ and $t'$
have an {\em $S$-clash} if they have different constructor symbols of $S$ at the same
position.

\subsection{The Proof Calculus for {\it CRWL}}
\label{sec:cal_crwl}

A {\it CRWL}-program $\mathcal{P}$ is a
finite
set of conditional rewrite rules of the form:
\begin{center}
  $\underbrace{f(t_1,...,t_n)}_{head}\eqr \underbrace{e}_{body}\Leftarrow
  \underbrace{C_1,...,C_m}_{condition}$
\end{center}

\noindent where $f\in FS^n$, and fulfilling the following conditions:
\begin{itemize}
\item[$\bullet$]  $(t_1,...,t_n)$ is a linear tuple (each variable in it occurs
  only once) with $t_1,...,t_n\in CTerm$;
\item[$\bullet$] $e\in Term$;
\item[$\bullet$] each $C_i$ is a constraint of the form $e'\con e''$ ({\em
    joinability}) or $e'\div e''$ ({\em divergence}) where $e',e''\in Term$;
\item[$\bullet$] {\em extra variables} are not allowed, i.e., all the variables
  appearing in the body $e$ and the condition $\overline{C}$ must also appear in
  the head $f(\overline{t})$ ($var(e)\cup var(\overline{C})\subseteq
  var(\overline{t})$).
This condition is not required in \cite{GHL96,GHL99};
see the end of this section for a discussion of this issue.
\end{itemize}

The reading of the rule is: $f(t_1,...,t_n)$ reduces to $e$ if the conditions
$C_1,...,C_n$ are satisfied. We write $\mathcal{P}_f$ for the set of defining
rules of $f$ in $\mathcal{P}$.

Given a program $\mathcal{P}$, the proof calculus for {\it CRWL} can
derive from it three kinds of statements:
\begin{itemize}
\item[$\bullet$] {\em Reduction or approximation statements}: $e\to t$, with
  $e\in Term_{\bot}$ and $t\in CTerm_{\bot}$. The intended meaning of such
  statement is that $e$ can be reduced to $t$, where reduction may be done by
  applying rewriting rules of $\mathcal{P}$ or by replacing subterms of $e$ by
  $\bot$. If $e \to t$ can be derived, $t$ represents one of the possible values
  of the denotation of $e$.
\item[$\bullet$] {\em Joinability statements}: $e\con e'$, with $e,e'\in Term_{\bot}$. The
  intended meaning in this case is that $e$ and $e'$ can be both reduced to some
  common totally defined value, that is, we can prove $e\to t$ and $e'\to t$ for
  some $t\in CTerm$.
\item[$\bullet$] {\em Divergence statements}: $e\div e'$, with $e,e'\in Term_{\bot}$. The
  intended meaning now is that $e$ and $e'$ can be reduced to some (possibly
  partial) c-terms $t$ and $t'$ having a $DC$-clash.
In \cite{GHL96,GHL99} divergence conditions are not considered. They
  have been incorporated to {\it CRWL} in \citeN{LS99b} as a useful and
  expressive resource for programming that is implemented in the system \toy.
\end{itemize}

When using function rules to derive statements, we will need to use what are
called {\em c-instances} of such rules. The set of c-instances of a program rule
$R$ is defined as:
\begin{center}
  $[R]_{\bot}=\{R\theta | \theta\in CSubst_{\bot}\}$
\end{center}
Parameter passing in function calls will be expressed by means of these
c-instances in the proof calculus.

\begin{table}[tbp]
  \caption{Rules for {\it CRWL}-provability}
  \begin{center}
    \begin{tabular}{l}
      \hline\hline\\

      (1)$\ $
      $\frac{}{\mbox{$
          \begin{array}{c}
            \\[-0.4cm]
            e\to\bot\\
            \\[-0.4cm]
          \end{array}$}}$\\[0.4cm]

      (2)$\ $
      $\frac{}{\mbox{$
          \begin{array}{c}
            \\[-0.4cm]
            X\to X\\
            \\[-0.4cm]
          \end{array}$}}$
      $\qquad  X\in \var$ \\[0.4cm]

      (3)$\ $
      $\frac{\mbox{$
          \begin{array}{c}
            \\[-0.4cm]
            e_1\to t_1,...,e_n\to t_n\\
            \\[-0.4cm]
          \end{array}$}}
      {\mbox{$
          \begin{array}{c}
            \\[-0.4cm]
            c(e_1,...,e_n)\to c(t_1,...,t_n)\\
            \\[-0.4cm]
          \end{array}$}}$
      $\qquad c\in DC^n,\quad t_i\in CTerm_{\bot}$\\[0.4cm]

      (4)$\ $
      $\frac{\mbox{$
          \begin{array}{c}
            \\[-0.4cm]
            e_1\to s_1,...,e_n\to s_n\quad C\quad e\to t\\
            \\[-0.4cm]
          \end{array}$}}
      {\mbox{$
          \begin{array}{c}
            \\[-0.4cm]
            f(e_1,...,e_n)\to t\\
            \\[-0.4cm]
          \end{array}$}}$
        $\qquad
        \begin{array}{l}
          \textrm{if $t\not\equiv\bot, R\in\mathcal{P}_f$}\\
          (f(s_1,...,s_n)\eqr e \Leftarrow C)\in [R]_{\bot}
        \end{array}$\\[0.4cm]

      (5)$\ $
      $\frac{\mbox{$
          \begin{array}{c}
            \\[-0.4cm]
            e\to t\quad e'\to t\\
            \\[-0.4cm]
          \end{array}$}}
      {\mbox{$
          \begin{array}{c}
            \\[-0.4cm]
            e \con e'\\
            \\[-0.4cm]
          \end{array}$}}$
      $\qquad$ if $t\in CTerm$\\[0.4cm]

      (6)$\ $
      $\frac{\mbox{$
          \begin{array}{c}
            \\[-0.4cm]
            e\to t\quad e'\to t'\\
            \\[-0.4cm]
          \end{array}$}}
      {\mbox{$
          \begin{array}{c}
            \\[-0.4cm]
            e \div e'\\
            \\[-0.4cm]
          \end{array}$}}$
      $\qquad$ if $t,t'\in CTerm_\bot$ and have a $DC-$clash\\[0.4cm]

      \hline\hline
    \end{tabular}
    \label{tab:crwl}
  \end{center}
\end{table}

Table \ref{tab:crwl} shows the proof calculus for {\it CRWL}. We write
$\mathcal{P}\conscrwl \varphi$ for expressing that the statement $\varphi$ is
provable from the program $\mathcal{P}$ with respect to this calculus.
The rule (4) allows
to use c-instances of program rules to prove approximations. These c-instances
may contain $\bot$ and by rule (1) any expression can be reduced to $\bot$.
 This reflects a non-strict semantics. A variable $X$
can only be approximated by itself (rule 2) and by $\bot$ (rule 1), so
a variable is similar to a constant in derivations with this calculus.
Nevertheless, when using function rules of the program a variable of such rule
can take any value by taking the appropriate c-instance. The rule (3) is
for term decomposition and rules (5) and (6) corresponds to the definition of $\con$
and $\div$ respectively.

A distinguished feature of {\it CRWL} is that functions can be
{\em non-deterministic}. For example, assuming the constructors $z$
(zero) and $s$ (successor) for natural numbers, a non-deterministic
function {\em coin} for expressing the possible results of throwing a coin can
defined by the rules:
\begin{center}
  $
  \begin{array}{l}
    coin \eqr z\\
    coin \eqr s(z)
  \end{array}$
\end{center}

It is not difficult to see that the previous calculus can derive the statement
$coin\eqr z$ and also $coin\eqr s(z)$. The use of c-instances in rule $(4)$
instead of general instances corresponds to {\em call time choice} semantics for
non-determinism (see \cite{GHL99}). As an example, in addition to {\em coin}
consider the functions {\em add} and {\em double} defined as:
\begin{center}
  $
  \begin{array}{ll}
    \begin{array}{l}
      add(z,Y)\eqr Y\\
      add(s(X),Y)\eqr s(add(X,Y))
    \end{array}
    \qquad &\qquad
    \begin{array}{l}
      double(X)\eqr add(X,X)
    \end{array}
  \end{array}$
\end{center}

It is possible to build a {\it CRWL}-proof for the statement $double(coin)\to z$
and also for $double(coin)\to s(s(z))$, but not for $double(coin)\to s(z)$. As
an example of derivation, we show a derivation for $double(coin)\to z$; at each
step we indicate by a number on the left the rule of the calculus applied:

\medskip
\begin{center}
  $\cd$
\end{center}
\medskip

Observe that $\div$ is not the logical negation of $\con$. They are not even
incompatible: due to non-determinism, two expressions $e,e'$ can satisfy both
$e\con e'$ and $e\div e'$ (although this cannot happen if $e,e'$ are c-terms).
In the `coin' example, we can derive both $coin\con z$ and $coin\div z$.

\medskip

The {\em denotation} of an expression $e$ can be defined as the set of c-terms to
which $e$ can be reduced according to this calculus:
\begin{center}
  $[\! [ e ]\! ]=\{ t\in CTerm_{\bot} | \mathcal{P}\conscrwl e\to t\}$
\end{center}
For instance, $[\! [ coin ]\! ]=\{\bot, z, s(\bot),s(z)\}$.

To end our presentation of the {\it CRWL} framework we
discuss the issue of extra variables (variables not appearing in left hand sides of function rules),
which are allowed in \cite{GHL96,GHL99}, but not in this paper. This
 is not {\em as} restrictive as it could appear:
function nesting can replace the use (typical of logic programming) of variables
  as repositories of intermediate values, and in many other cases where extra
  variables represent unknown values to be computed by search, they can be successfully
replaced by non-deterministic functions able to compute candidates for such unknown values.
A concrete example is given by the function {\it next} in example \ref{grafo}.
More examples can be found in \cite{GHL99,Toy}.

The only extra variable we have used in Sect. \ref{sec:interest} is $Pos'$ in the definition
\begin{center}
 $winMove(Pos) \eqr Pos' \Leftarrow Pos' \con move(Pos), {\it fails}(winMove(Pos')) \con true $
\end{center}
of example \ref{nim}. It can be removed by introducing an auxiliary function:
\begin{center}
$\begin{array}{l}
winMove(Pos) \eqr aux(move(Pos))\\
aux(Pos)  \eqr Pos \Leftarrow Pos \con Pos, {\it fails}(winMove(Pos)) \con true
\end{array}$
\end{center}
The effect of the condition $Pos \con Pos$ it to compute a normal form for
$Pos$, which is required in this case to avoid a diverging computation for
{\em winMove(Pos)}.

\section{The {\it CRWLF} Framework}
\label{sec:crwl_framework}

We now address the problem of failure in {\it CRWL}.
Our primary interest is to obtain a calculus able to prove that
a given expression fails to be reduced. Since reduction corresponds
in {\it CRWL} to approximation statements $e\to t$, we can reformulate our aim
more precisely: we look for a calculus able to prove that a given
expression $e$ has no possible reduction (other than the trivial $e\to\bot$)
in {\it CRWL}, i.e., $[\! [ e ]\! ] = \{\bot\}$.

Of course, we cannot expect to achieve that with full generality
since, in particular, the reason for having $[\! [ e ]\! ] =
\{\bot\}$ can be non-termination of the program as rewrite system, a property
which is uncomputable. Instead, we look for a
suitable computable approximation
%\footnote{We cannot even speak of {\em the best} computable approximation}
to the property $[\! [ e ]\! ] = \{\bot\}$, corresponding to cases where failure of reduction is due
to `finite' reasons, which can be constructively detected and managed.
% We can speak informally of {\em finite failure}.

Previous to the formal presentation of the calculus, which will be called
\mbox{{\it CRWLF}} (for `{\it CRWL} with failure')  we give several simple
examples for a preliminary understanding of some key aspects of it,
and the reasons underlying some of its technicalities.

\subsection{Some Illustrative Examples}
\label{sec:examples}

Consider the following functions, in addition to $coin$, defined in Sect. \ref{sec:cal_crwl}:
\begin{center}
$\begin{array}{lll}
f(z)\eqr f(z)\qquad g(s(s(X)))\eqr z\qquad
    & \begin{array}{l}
      h \eqr s(z)\\
      h \eqr s(h)
    \end{array}\qquad
    & k(X) \eqr  z \Leftarrow X\con s(z)
\end{array}$
\end{center}

We discuss several situations involving failure with this program:
\begin{itemize}
\item[$\bullet$] The expressions $f(z)$ and $f(s(z))$ fail to be reduced, but for quite different
reasons. In the first case $f(z)$ does not terminate. The only possible proof
accordingly to {\it CRWL} is $f(z)\to\bot$ (by rule $1$); any attempt to prove
$f(z)\to t$ with $t\not =\bot$ would produce an `infinite derivation'. In the second
case, the only possible derivation is again $f(s(z))\to\bot$, but if we try to prove
$f(s(z))\to t$ with $t\not =\bot$ we have a kind of `{\em finite failure}': rule $4$ needs
to solve the parameter passing $s(z)\to z$, that could be finitely checked as
failed, since no rule of the {\it CRWL}-calculus is applicable.
The {\it CRWLF}-calculus does not prove non-termination of
$f(z)$, but will be able to detect and manage the failure for $f(s(z))$. In
fact it will be able to perform a {\em constructive proof} of this failure.
\item[$\bullet$] Consider now the expression $g(coin)$. Again, the only possible reduction is
$g(coin)\to\bot$ and it is intuitively clear that this is another case of finite
failure. But this failure is not as simple as in the previous example for
$f(s(z))$: in this case the two possible reductions for $coin$ to defined values are $coin\to
z$ and $coin\to s(z)$. Both of $z$ and $ s(z)$ fail to match the pattern $s(s(X))$
in the rule for $g$, but none of them
can be used separately to detect the failure of $g(coin)$.
A suitable idea is to collect the set of defined values to which
a given expression can be reduced. In the case of $coin$
that set is $\{ z,s(z)\}$.
The fact  that $\cal C$ is the collected set of values
of $e$ is expressed in {\it CRWLF} by means of the
statement $e \rec {\cal C}$. In our example, {\it CRWLF} will prove
$coin \rec \{ z,s(z)\}$.  Statements $e \rec {\cal C}$ generalize the approximation
statements $e \to t$ of {\it CRWL}, and in fact can replace them. Thus,
{\it CRWLF} will not need to use explicit $e \to t$ statements.
\item[$\bullet$] How far should we go when collecting values? The idea of collecting
all values (and to have them completely evaluated)
works fine in the previous example, but there are problems
when the collection is infinite.
For example, according to its definition above, the expression
$h$  can be reduced to any positive natural number, so the corresponding set
would be $H = \{s(z),s(s(z)),s(s(s(z))), ...\}$. Then, what if we try to reduce the
expression $f(h)$? From an intuitive
point of view it is clear that the value $z$ will not appear in $H$,
because all its elements have the form $s(...)$.
The partial value $\{ s(\bot)\}$ is a common approximation to all the elements of $H$.
Here we can understand $\bot$ as an {\em
  incomplete information}: we know that all the  values for $h$ are
successor of `something', and this implies that they cannot be $z$,
which suffices for proving the failure of $f(h)$.
The {\it CRWLF}-calculus will be able to prove the
statement $h \rec \{ s(\bot)\}$, and we say that $\{ s(\bot)\}$
is a {\em Sufficient Approximation Set} ({\it SAS}) for $h$.

In general, an expression will have multiple {\it SAS}'s. Any
expression has $\{\bot\}$ as its simplest {\it SAS}.  And, for example, the expression $h$ has
an infinite number of {\it SAS}'s:
$\{\bot\}$, $\{s(\bot)\}$, $\{ s(z),s(s(\bot))\}$,...
The {\it SAS}'s obtained by the calculus for $coin$ are
$\{\bot\}$, $\{\bot,s(\bot)\}$,$\{\bot,s(z)\}$, $\{z,\bot\}$,  $\{z,s(\bot)\}$ and
$\{z,s(z)\}$.
The {\it CRWLF}-calculus
provides appropriate rules for working with {\it SAS}'s. The derivation steps
will be guided by these {\it SAS}'s in the same sense that {\it CRWL} is guided by
approximation statements.

\item[$\bullet$] Failure of reduction is due in many cases to failure in proving the
conditions in the program rules. The calculus must be able to prove those failures.
Consider for instance the expression $k(z)$. In this case we would try to use the
c-instance $k(z)\eqr z\Leftarrow z\con s(z)$ that allows to perform parameter
passing. But the condition
$z\con s(z)$ is clearly not provable, so $k(z)$ must fail. For
achieving it we must be able to give a proof for {\em
  `$z \con s(z)$ cannot be proved with respect to {\it CRWL}'}. For this purpose
we introduce a new constraint $e\nocon e'$ that will be true if we can build a
{\em proof of non-provability} for $e\con e'$. In our case, $z\nocon s(z)$ is
clear because of the clash of constructors. In general the proof for a
constraint $e\nocon e'$ will be guided by the corresponding {\it SAS}'s for $e$ and
$e'$ as we will see in the next section. As our initial {\it CRWL} framework also
allows constraints of the form $e\div e'$, we need also another constraint
$\nodiv$ for expressing `failure of $\div$'.

\item[$\bullet$] There is another important question to justify: we use an explicit
representation for failure by means of the new constant symbol $\fail$. Let us
examine some examples involving failures. First, consider the expression
$g(s(f(s(z))))$; for reducing it we would need to do parameter passing,
i.e., matching $s(f(s(z)))$ with some c-instance of the pattern
$s(s(X))$ of the definition of $g$. As $f(s(z))$ fails to be reduced the
parameter passing must also fail. If we take $\{\bot\}$ as an {\it SAS} for $f(s(z))$
we have not enough information for detecting the failure (nothing can be said
about the matching of $s(s(X))$ and
$s(\bot)$). But if we take $\{\fail\}$ as an {\it SAS} for $f(s(z))$, this provides
enough information to ensure that $s(\fail)$ cannot match any
c-instance of the pattern $s(s(X))$. Notice that we allow the value
$\fail$ to appear inside the term $s(\fail)$. One could think that the
information $s(\fail)$ is essentially the same of $\fail$ (for instance,
$\fail$ also fails to match any c-instance of $s(s(X))$), but this
is not true in general. For instance, the
expression $g(s(s(f(s(z)))))$ is reducible  to $z$. But if we take the {\it SAS}
$\{\fail\}$ for $f(s(z))$ and we identify the expression $s(s(f(s(z))))$ with
$\fail$, matching with the rule for $g$ would not succeed, and the
reduction of $g(s(s(f(s(z)))))$ would fail.

\end{itemize}

\medskip

We can now proceed with the formal presentation of the {\it CRWLF}-calculus.

\subsection{Technical Preliminaries}
\label{sec:prelim_crwlf}

For dealing with failure we consider two new syntactical elements in {\it
  CRWLF}: a function ${\it fails}$ and a constant $\fail$. The first one is
directly included into the signature, so we consider $\Sigma =DC\cup
FS\cup\{{\it fails}\}$, where $DC$ and $FS$ are sets of {\it constructor} symbols
and (user-defined) {\it functions} respectively. This symbol, ${\it fails}$,
stands for a predefined function whose intuitive meaning is:

\begin{center}
  ${\it fails}(e)::=
  \left\{ \begin{array}{ll}
    {\it true} & \textrm{if $e$ fails to be reduced to hnf}\\
    {\it false} & \textrm{otherwise}
  \end{array}\right.$
\end{center}

The boolean constants ${\it true}$ and ${\it false}$ must belong to $DC$, as they are
needed to define the function ${\it fails}$. The formal interpretation of this
function will be defined by specific rules at the level of the proof-calculus
(Table \ref{tab:crwlf}).

The second syntactical element, the constant $\fail$, is introduced as an
extension of the signature (as it was the element $\bot$ in {\it CRWL}). So we
use the extended signature $\Sigma_{\bot,\fail}=\Sigma\cup\{\bot,\fail\}$. We do not
include it directly in the signature $\Sigma$ because its role is to
express failure of reduction and it is not allowed to appear explicitly in a
program. In the case of the function ${\it fails}$ we want to allow to use it in
programs as we have seen in the examples of Sect. \ref{sec:interest}.

The sets $Term_{\bot,\fail}, CTerm_{\bot,\fail}$ are defined in the natural
way, and also the set of substitutions $CSubst_{\bot,\fail}=\{\theta:\var\to
CTerm_{\bot,\fail}\}$.

A natural {\em approximation ordering} $\sqsubseteq$ over $Term_{\bot,\fail}$
can be defined as the least partial ordering over $Term_{\bot,\fail}$ satisfying
the following properties:

  \begin{itemize}
  \item[$\bullet$] $\bot \sqsubseteq e$ for all $e\in Term_{\bot,\fail}$,
  \item[$\bullet$] $h(e_1,...,e_n)\sqsubseteq h(e'_1,...,e'_n)$, if $e_i\sqsubseteq e'_i$
    for all $i\in\{1,...,n\}$, $h\in DC\cup FS\cup\{{\it fails}\}$
  \end{itemize}

The intended meaning of $e \sqsubseteq e'$ is that $e$ is less defined or has less
information than $e'$.  Two expressions $e,e'\in Term_{\bot,\fail}$ are
{\em consistent} if they can be refined to obtain the same information,
i.e., if there exists $e''\in
Term_{\bot,\fail}$ such that $e\sqsubseteq e''$ and $e'\sqsubseteq e''$.  

Notice that the only relations satisfied by $\fail$ are $\bot \sqsubseteq \fail$ and 
$\fail \sqsubseteq \fail$. In particular,  $\fail$ is maximal. This is
reasonable, since $\fail$ represents `failure of reduction' and 
this gives no further refinable information about the
result of the evaluation of an expression. 
This contrasts with the status given to failure in \cite{Moreno96ELP},
where $\fail$ is chosen to verify $\fail \sqsubseteq t$ for any $t$ different
from $\bot$.

We will frequently use the following notation:
given $e\in Term_{\bot,\fail}$, $\hat{e}$ stands for
the result of replacing by $\bot$ all the occurrences of $\fail$ in $e$
(notice that $\hat{e}\in Term_{\bot}$, and 
$e= \hat{e}$ iff $e \in Term_{\bot}$).

\subsection{The Proof Calculus for {\it CRWLF}}
\label{sec:cal_crwlf}

Programs in {\it CRWLF} are sets of rules with the same form as in {\it CRWL},
but now they can make use of the function ${\it fails}$ in the body and in the
condition part, i.e., {\it CRWLF} extends the class of programs of {\it
  CRWL} by allowing the use of {\it fails} in programs. On the other hand, in {\it
  CRWLF}  five kinds of statements can be deduced: 
\begin{itemize}
\item[$\bullet$] $e \rec {\cal C}$, intended to mean `{\em ${\cal C}$ is an {\it SAS} for $e$}'.
\item[$\bullet$] $e \con e'$, $e \div e'$, with the same intended meaning as in {\it CRWL}.
\item[$\bullet$] $e \nocon e'$, $e\nodiv e'$, intended to mean failure of $e
  \con e'$ and $e \div e'$ respectively.
\end{itemize}

We will sometimes speak of $\con, \div, \nocon, \nodiv$ as `constraints', and
use the symbol $\diamondsuit$ to refer to any of them. The constraints
$\nocon$ and $\con$ are called the {\em complementary} of each other; 
the same holds for $\nodiv$ and $\div$, and we write 
$\widetilde{\diamondsuit}$ for the complementary of $\diamondsuit$.

When proving a constraint $e\diamondsuit e'$ the calculus {\it CRWLF} will evaluate an
{\it SAS} for the expressions $e$ and $e'$. These {\it SAS}'s will consist 
of c-terms from $CTerm_{\bot,\fail}$, and provability of the constraint $e\diamondsuit e'$
depends on certain syntactic (hence decidable) relations between those c-terms.
Actually, the constraints $\con$, $\div$, $\nocon$ and $\nodiv$ can be seen
as  the result of generalizing to expressions the relations
$\downarrow,\uparrow,\not\downarrow$ and $\not\uparrow$ on c-terms,  
which we define now.  

\begin{definition}[Relations over $CTerm_{\bot,\mbox{\textnormal \fail}}$] 
\label{def:arrows}
  \begin{itemize}
  \item[$\bullet$] $t\downarrow t'\Leftrightarrow_{def} t=t', t\in CTerm$
  \item[$\bullet$] $t\uparrow t' \Leftrightarrow_{def} t$ and $t'$ have a $DC$-clash
  \item[$\bullet$] $t\not\downarrow t'\Leftrightarrow_{def} t$ or $t'$ contain $\fail$ as
    subterm, or they have a $DC$-clash
  \item[$\bullet$] $\not\uparrow$ is defined as the least symmetric relation over
    $CTerm_{\bot,\fail}$ satisfying:
    \begin{itemize}
    \item[$i)$] $X\not\uparrow X$, for all $X\in\var$
    \item[$ii)$] $\fail\not\uparrow t$, for all $t\in CTerm_{\bot,\fail}$
    \item[$iii)$] if $t_1\not\uparrow t'_1,...,t_n\not\uparrow t'_n$ then
      $c(t_1,...,t_n)\not\uparrow c(t'_1,...,t'_n)$, for $c\in DC^n$
    \end{itemize}
  \end{itemize}
\end{definition}

\medskip

The relations $\downarrow$ and $\uparrow$  do not take
into account the presence of $\fail$, which behaves in this case as $\bot$. The 
relation $\downarrow$ is
{\em strict} equality, i.e., equality restricted to total c-terms. It is the notion of equality
used in lazy functional or functional-logic languages as the suitable
approximation to `true' equality ($=$) over $CTerm_{\bot}$.  The relation 
$\uparrow$ is a suitable approximation to `$\neg =$', and hence to `$\neg \downarrow$'
(where $\neg$ stands for logical negation). The relation $\not\downarrow$ is also an approximation
to `$\neg \downarrow$', but in this case using failure information ($\not\downarrow$ can be read
as `$\downarrow$ fails'). Notice that
$\not\downarrow$ does not imply `$\neg =$' anymore (we have, for instance,  $\fail
\not\downarrow \fail$). Similarly, $\not\uparrow$ is also an approximation to `$\neg
\uparrow$' which can be read as `$\uparrow$ fails'.

The following proposition reflects these and more good properties of
$\downarrow, \uparrow, \not\downarrow, \not\uparrow$.

\begin{proposition}
\label{prop:1}
The relations $\downarrow, \uparrow,\not\downarrow, \not\uparrow$ satisfy
  \begin{itemize}
  \item[$a)$] For all $t,t',s,s'\in CTerm_{\bot,\fail}$
    \begin{itemize}
     \item[$i)$] $t\downarrow t'\Leftrightarrow \hat{t}\downarrow\hat{t}'$
        and  $t\uparrow t'\Leftrightarrow \hat{t}\uparrow\hat{t}'$ 
    \item[$ii)$] $t\uparrow t'\Rightarrow t\not\downarrow t'\Rightarrow \neg
(t\downarrow t')$
   \item[$iii)$] $t\downarrow t'\Rightarrow t\not\uparrow t'\Rightarrow \neg
(t\uparrow t')$
    \end{itemize}
    
  \item[$b)$]%{\bf Monotonicity:}
    $\downarrow, \uparrow,
    \not\downarrow, \not\uparrow$ are monotonic, i.e., if $t\sqsubseteq s$ and
    $t'\sqsubseteq s'$ then: $t \Re  t'\Rightarrow s \Re s'$, where
    $\Re\in\{\downarrow, \uparrow, \not\downarrow, \not\uparrow\}$. Furthermore
 $\not\downarrow_G$ and $\not\uparrow_G$ are the greatest monotonic
approximations to
      $\neg\downarrow_G$ and $\neg\uparrow_G$ respectively, where $\Re_G$ is
 the restriction of $\Re$ to the set of ground (i.e., without variables) c-terms
from $CTerm_{\bot,\fail}$.

  \item[$c)$] %{\bf Substitutions:}
  $\downarrow$ and $\not\uparrow$ are
    closed  under substitutions from $CSubst$;  $\not\downarrow$ and $\uparrow$
    are closed under substitutions from $CSubst_{\bot,\fail}$
\end{itemize}
\end{proposition}

\begin{proof*}
  We prove each property separately:  
  \begin{itemize}
  \item[$a)$]
    \begin{itemize}
    \item[$i)$]
      \begin{itemize}
      \item[$\bullet$] $t\downarrow t'\Leftrightarrow \hat{t}\downarrow\hat{t}'$: two
      terms satisfying the relation $\downarrow$ cannot contain $\bot$ neither
      $\fail$. %, what means that they remain identical under the hat, so
      Hence $t=\hat{t}$ and $t'=\hat{t}'$, and the equivalence is trivial.

      \item[$\bullet$] $t\uparrow t'\Leftrightarrow \hat{t}\uparrow\hat{t}'$: the relation
      $\uparrow$ is satisfied when the terms have a $DC$-clash at some position $p$;
      since $t$ and $\hat{t}$
      ($t'$ and $\hat{t}'$ resp.) have the same constructor symbols at the
same positions, the equivalence is clear.
      \end{itemize}

    \item[$ii)$] The implication $t\uparrow t'\Rightarrow t\not\downarrow t'$ is
      clear from definitions of $\uparrow$ and $\not\downarrow$. For
      $t\not\downarrow t'\Rightarrow \neg (t\downarrow t')$: if $t\not\downarrow
      t'$ then either $\fail$ appears in $t$ or $t'$, or $t$
      and $t'$ have a $DC$-clash. In both cases $t\downarrow t'$ does not hold.

    \item[$iii)$] For $t\downarrow t'\Rightarrow t\not\uparrow t'$: if
      $t\downarrow t'$ then $t=t'$ with $t\in CTerm$ and we have $t\not\uparrow
      t'$ by applying repeatedly $i)$ and $iii)$ of the definition 
      of $\not\uparrow$. For $t\not\uparrow t'\Rightarrow \neg (t\uparrow t')$
      let us assume $t\not\uparrow t'$ and proceed by induction on the depth
      $d$ of $t$:
        
        $\underline{d=0}$: if $t=\bot$ or $t =\fail$ then $t$ and $t'$ cannot
        have any $DC$-clash and then $t\uparrow t'$ is not true. If $t = X$ or $t
        = c\in DC^0$ then $t\not\uparrow t'$ implies that $t' = \fail$ or $t'
        = t$; therefore $t$ and $t'$ cannot have any
        $DC$-clash and $t\uparrow t'$ is not true.

      $\underline{d \Rightarrow d+1}$: if $t=c(t_1,...,t_n)$,  then either  $t'=\fail$ and  $t\uparrow
      t'$ is not true, or $t'=c(t'_1,...,t'_n)$ with $t_i\not\uparrow
      t'_i$ for all $i\in\{1,...,n\}$; in this case, by i.h. there is not a pair
      $(t_i,t'_i)$ with a $DC$-clash, so neither $t$ and $t'$
      have $DC$-clashes, and therefore $t\uparrow t'$ is not true.

    \end{itemize}

  \item[$b)$] We prove monotonicity for each relation:
    \begin{itemize}
    \item[$\bullet$] For $\downarrow$: by definition of $\downarrow$, if  $t\downarrow t'$ then $t,t'\in
      CTerm$ (they are maximal with respect to $\sqsubseteq$), hence  $s=t$
      and $s'=t'$ and then $s\downarrow s'$.
    \item[$\bullet$] For $\uparrow$: if $t\uparrow t'$ then $t$ and $t'$ have a $DC$-clash
      at some position. As 
      $t\sqsubseteq s$ and $t'\sqsubseteq s'$, then $s$ 
      and $s'$ will have the same $DC$-clash at the same position, so $s\uparrow s'$.
    \item[$\bullet$] For $\not\downarrow$: if $t$ and $t'$ have a $DC$-clash, $s$ and $s'$ will
      contain the same $DC$-clash, as in $ii)$. If one of them has $\fail$ as subterm,
      by definition of $\sqsubseteq$ it is clear that $s$ or $s'$ will also contain
      $\fail$, so $s\not\downarrow s'$.
    \item[$\bullet$] For $\not\uparrow$: Here we proceed by induction on the depth $d$ of the term $t$:
      
      $\underline{d=0}$: let us check the possibilities for $t$. If $t=X$ or $t=c\in
      DC^0$, then $t\not\uparrow t'$ implies  $t'=t$ or $t'=\fail$;
    since $t,t'$ are maximal with respect to $\sqsubseteq$, then $s=t$
       and $s'=t'$, so we will also have $s\not\uparrow s'$. If $t=\fail$ then
      $s=\fail$ and then it is clear that $s\not\uparrow s'$. If $t=\bot$ then
      $t'=\fail=s'$ and it is clear that $s\not\uparrow s'$.
      
      $\underline{d \Rightarrow d+1}$: in this case  $t=c(t_1,...,t_n)$ and then either
      $t'=\fail$, what implies $s'=\fail$ and then $s\not\uparrow s'$, or  $t'=c(t'_1,...,t'_n)$ with
      $t_i\not\uparrow t'_i$ for all $i\in\{1,...,n\}$. From $t\sqsubseteq s$ and
      $t'\sqsubseteq s'$ it follows that $s=c(s_1,...,s_n)$ and $s'=c(s'_1,...,s'_n)$,
      and by i.h. we have $s_i\not\uparrow s'_i$ for all $i\in\{ 1,...,n\}$, what
      implies $s\not\uparrow s'$.
    \end{itemize}
    
    \medskip

    Now we prove that $\not\downarrow_G$ and $\not\uparrow_G$ are the
    greatest monotonic approximations to $\neg\downarrow_G$ and $\neg\uparrow_G$
    respectively. 
We note by $GCTerm_{\bot,\fail}$ the set of all ground $t \in CTerm_{\bot,\fail}$.
   \begin{itemize}
    \item[$\bullet$] For $\not\downarrow_G$, assume that a relation
      $R\subseteq (GCTerm_{\bot,\fail}\times GCTerm_{\bot,\fail})$ verifies
      \begin{center}
        $\begin{array}{l}
            t R t'\Rightarrow \neg(t\downarrow_G t')\\
            t\sqsubset s, t'\sqsubseteq s', t R t'\Rightarrow s R s'
          \end{array}$
      \end{center}
We must prove that $R$ is included in $\not\downarrow_G$, that is: $(t R t'\Rightarrow t\not\downarrow_G t')$, for any $t,t' \in GCTerm_{\bot,\fail}$.
We reason by contradiction.
      Assume $t R t'$ and $\neg(t\not\downarrow_G t')$. Then, by definition of $\not\downarrow_G$,
       $t$ and $t'$ do not contain $\fail$  and do not
      have a $DC$-clash.
      Then either $t = t'$,
      or $t$ and $t'$ differ because at some positions one of them has $\bot$ while the other has not.
      In both cases it is easy to see that  there  exists  $s\in GCTerm$
      (totally defined) such that $t\sqsubseteq s$ and $t'\sqsubseteq s$. By
      monotonicity of $R$ we have $s R s$ what implies $\neg(s\downarrow_G s)$, what is
      a contradiction, since $s\in CTerm$.

    \item[$\bullet$] For $\not\uparrow_G$ we proceed in a similar way as in the previous point: assuming
     that  $R\subseteq (GCTerm_{\bot,\fail}\times GCTerm_{\bot,\fail})$ verifies 

      \begin{center}
        $\begin{array}{l}
            t R t'\Rightarrow \neg(t\uparrow_G t')\\
            t\sqsubset s, t'\sqsubseteq s', t R t'\Rightarrow s R s'
          \end{array}$
      \end{center}            
we must prove $(t R t'\Rightarrow t\not\uparrow_G t')$. But 
      if $t R t'$ then $\neg (t\uparrow_G t')$, so $t$ and $t'$ cannot have any
      $DC$-clash. They could contain $\fail$ as subterm but then, by $ii)$ and $iii)$ of the
      definition of $\not\uparrow$, we will have $t\not\uparrow_G t'$.
    \end{itemize}

  \item[$c)$] The property is clear for $\downarrow$: if we
    replace in a c-term all the 
    occurrences of a variable by a totally defined c-term, we will obtain a totally
    defined c-term. For $\uparrow$, such substitution preserves 
    the $DC$-clash of the original c-terms.

    For $\not\downarrow$, if some of the original c-terms had $\fail$ as a subterm,
    the substitution preserves this occurrence of $\fail$. On the other hand, if they
    had a $DC$-clash, then it is clear that this clash will also be present under the
    substitution.  
    
    For $\not\uparrow$, suppose $t\not\uparrow t'$ and $\theta\in
    CSusbt_{\bot,\fail}$; we proceed by induction on the depth $d$ of the 
    term $t$:
    
    $\underline{d=0}$: if $t=\fail$, then $t\theta=\fail$ and it is clear that
    $t\theta\not\uparrow t'\theta$. For the cases $t=X$ and $t=c\in DC^0$ we have two
    possibilities for $t'$: $t'=\fail$ or $t'=t$; if $t'=\fail$ the result is
    clear. If we have $t=t'=X$ it is not difficult to prove that 
    $X\theta\not\uparrow X\theta$ by applying repeatedly $i)$ and $iii)$ of
    definition of $\not\uparrow$. The last case, if $t=t'=c\in DC^0$ is trivial
    because $\theta$ does not change the terms.

    $\underline{d \Rightarrow d+1}$: in this case  $t=c(t_1,...,t_n)$. If $t'=\fail$ the proof is as
    in the base case, otherwise $t'=c(t'_1,...,t'_n)$ with $t_i\not\uparrow
    t'_i$ for all $i\in\{ 1,...,n\}$. By i.h. we have $t_i\theta\not\uparrow
    t'_i\theta$ and then, by $iii)$ of the 
    definition of $\not\uparrow$ we will have $t\theta\not\uparrow t'\theta$.
$\mathproofbox$
  \end{itemize}
\end{proof*}

\medskip

By {\it (b)}, we can
say that $\downarrow, \uparrow,\not\downarrow, \not\uparrow$ behave well with
respect to the information ordering: if they are true for some terms, they
remain true if we refine the information contained in the terms. Furthermore, 
{\it (b)} states that $\not\downarrow, \not\uparrow$ are defined `in the
best way' (at least for ground c-terms) as computable approximations to
$\neg\downarrow$ and $\neg\uparrow$. For c-terms with variables, we
must take care: for instance, given the constructor $z$, we have
$\neg(X\downarrow z)$, but
not $X\not\downarrow z$. Actually, to have $X\not\downarrow z$ would
violate a basic intuition about free variables in logical
statements: if the statement is true, it should be true for any
value (taken from an appropriate range) substituted for its free variables. 
The part {\it (c)} shows that the definitions of 
$\downarrow, \uparrow,\not\downarrow, \not\uparrow$ respect such
principle.
Propositions \ref{prop:2} and \ref{prop:3} of the next section show that
monotonicity and closure by substitutions are 
preserved when generalizing 
$\downarrow, \uparrow,\not\downarrow, \not\uparrow$ to
$\con, \div, \nocon, \nodiv$.

\medskip

We can present now the proof rules for the {\it CRWLF}-calculus, which are shown
in Table \ref{tab:crwlf}. The rules 6 and 7 use a generalized
notion of c-instances of a rule $R$: $[R]_{\bot,\fail} = \{ R\theta\  | \ \theta\in
CSubst_{\bot,\fail} \}$. We will use
the notation  $\mathcal{P}\conscrwlf\varphi$
($\mathcal{P}\not\conscrwlf\varphi$ resp.) for expressing that the statement
$\varphi$ is provable (is not provable resp.) with respect to the calculus {\it CRWLF} and the program
$\mathcal{P}$.
 {\it CRWLF}-derivations have a tree structure (see e.g. Example \ref{ex:5}); many results in the following
 sections use induction over the {\em size} of the derivation, i.e., the number of nodes in the derivation
 tree, which corresponds to the number of inference steps.

\begin{table}[h!]
  \caption{Rules for {\it CRWLF}-provability}
  \begin{center}
    \begin{tabular}{l}
      \hline\hline

      (1)$\ $ $\mifrac{}{e\rec \{ \bot\}}$\\[0.6cm]

      (2)$\ $ $\mifrac{}{X\rec \{ X\}}$ $\qquad X\in\var$\\[0.6cm]

      (3)$\ $ $\mifrac{e_1\rec \cb_1\quad ...\quad e_n\rec
        \cb_n}{c(e_1,...,e_n)\rec \{c(t_1 ,...,t_n)\ | \ \overline{t}\in
        \cb_1\times ...\times\cb_n \}}$ $\quad c\in DC^n\cup \{\fail\}$\\[0.6cm]

      (4)$\ $ $\mifrac{e_1\rec \cb_1\quad ...\quad e_n\rec \cb_n\quad...\quad
        f(\overline{t})\recr\cb_{R,\overline{t}}\quad ...}{f(e_1,...,e_n)\rec
 \bigcup_{R\in\mathcal{P}_f ,\overline{t}\in\cb_1\times
...\times\cb_n}
        \cb_{R,\overline{t}}}$  $\qquad f\in FS^n$\\[0.6cm]

      (5)$\ $ $\mifrac{}{f(\overline{t})\recr\{ \bot\}}$\\[0.6cm]

      (6)$\ $ $\mifrac{e\rec\cb\quad \overline{C}}{f(\overline{t})\recr\cb}$
      $\qquad$
      $(f(\overline{t})\eqr e \Leftarrow \overline{C})\in [ R
]_{\bot,\fail}$\\[0.6cm]

       (7)$\ $ $\mifrac{e_i\tilde{\diamondsuit} e'_i}{f(\overline{t})\recr \{
\fail\}}$
       $\qquad$
         $(f(\overline{t})\eqr e \Leftarrow ...,e_i\diamondsuit e'_i,...)\in
[ R
         ]_{\bot,\fail}$, where $i\in\{1,...,n\}$
         \\[0.6cm]

       (8)$\ $ $\mifrac{}{f(t_1,...,t_n)\recr \{ \fail\}}$
       \hspace{-2cm}
       \begin{tabular}{l}
         $R\equiv (f(s_1,...,s_n)\eqr  e \Leftarrow\overline{C}), t_i$ and
         $s_i$ have a\\
         $DC\cup\{\fail\}$-clash for some $i\in\{1,...,n\}$
       \end{tabular}\\[0.6cm]

       (9)$\ $ $\mifrac{e\rec\cb\quad e'\rec\cb'}{e \con e'}$
       $\qquad\exists t\in\cb, t'\in\cb'\  t\downarrow t'$\\[0.6cm]

       (10)$\ $ $\mifrac{e\rec\cb\quad e'\rec\cb'}{e\div e'}$
       $\qquad\exists t\in\cb, t'\in\cb'\ t\uparrow t'$\\[0.6cm]

       (11)$\ $ $\mifrac{e\rec\cb\quad e'\rec\cb'}{e\nocon e'}$
       $\qquad\forall t\in\cb, t'\in\cb'\ t\not\downarrow t'$\\[0.6cm]

      (12)$\ $ $\mifrac{e\rec\cb\quad e'\rec\cb'}{e\nodiv e'}$
      $\qquad\forall t\in\cb, t'\in\cb'\ t\not\uparrow t'$\\[0.6cm]

      (13)$\ $ $\mifrac{e\rec\{\fail\}}{{\it fails}(e)\rec\{{\it
true}\}}$\\[0.6cm]

      (14)$\ $ $\mifrac{e\rec\cb}{{\it fails}(e)\rec\{{\it false}\}}$
      $\qquad\exists t\in\cb, t\not =\bot, t\not =\fail$\\

      \hline\hline
    \end{tabular}
    \label{tab:crwlf}
  \end{center}
\end{table}

The first three rules are analogous to those of the {\it CRWL}-calculus, now dealing with
{\it SAS}'s instead of simple approximations (notice the cross product of {\it SAS}'s
in rule 3).
Rule 4 is a complex rule which requires some explanation to make clear its reading and,
more importantly, its decidability: to obtain an {\it SAS} $\cb$
for an expression $f(e_1,\ldots,e_n)$ (that is, to derive $f(e_1,\ldots,e_n) \rec \cb$)
we must first obtain  {\it SAS}'s  for $e_1,\ldots,e_n$ (that is, we must derive
$e_1 \rec \cb_1,\ldots, e_n  \rec \cb_n$);
then for each combination $\overline{t}$ of values in these {\it SAS}'s
(that is, for each $\overline{t} \in \cb_1 \times \ldots \times \cb_n$)
and each program rule $R$ for $f$, a part $\cb_{R,\overline{t}}$ of the whole {\it SAS} is produced;
the union of all these partial {\it SAS}'s constitutes the final {\em SAS} $\cb$ for $f(\overline{e})$.
Notice that since {\it SAS}'s are finite sets and programs are finite sets of rules, then there
is a finite number of $\cb_{R,\overline{t}}$ to be calculated in the premises of the rule,
and the union of all of them (the final calculated {\it SAS} in the rule) is again a finite set
\footnote{To be more precise, this reasoning would be the essential part of
an inductive proof of finiteness of {\it SAS}'s. But we do not think necessary to burden
the reader with such formality.}.

Rule 4
is quite different from rule 4 in {\it CRWL}, where we could use any c-instance of any
rule for $f$; here we need to consider simultaneously the contribution of
each rule to achieve `complete' information
about the values to which  the expression can be evaluated.
 We use the notation $f(\overline{t}) \rec_R \cb$ to indicate that
only the rule $R$ is used to produce $\cb$.  

Rules 5 to 8 consider all the possible ways in which a concrete rule $R$ can
contribute to the {\it SAS} of a call $f(\overline{t})$, where 
the arguments $\overline{t}$ are all in $CTerm_{\bot,\fail}$ (they come from the
evaluation of the arguments of a previous call $f(\overline{e})$). Rules 5
and 6 can be viewed as {\em positive} contributions. The first one obtains the
trivial {\it SAS} and 6 works if there is a c-instance of the rule $R$ with a head
identical to the head of the call (parameter passing); in this
case, if the constraints of this c-instance are provable, then the resulting
{\it SAS} is generated by the body of the c-instance. 
Rules 7 and 8 consider the {\em negative} or {\em failed}
contributions. Rule 7 applies when parameter passing can be done,
but it is possible to prove the complementary $e_i\tilde{\diamondsuit} e'_i$
of one of the
constraints $e_i \diamondsuit e'_i$ in the condition of the used c-instance. In this
case the constraint $e_i \diamondsuit e'_i$ (hence the whole condition in the c-instance) fails.
Finally, rule 8 considers the case in which parameter passing fails
because of a $DC\cup\{\fail\}$-clash between one of the arguments
in the call and the corresponding pattern in $R$.

We remark that for given $f(\overline{t})$ and $R$, the rule 5 and at most one
of rules 6 to 8 are applicable.  This fact, although intuitive, is far from
being trivial to prove  and constitutes in fact an important technical detail in
the proofs of the results in the next section.

Rules 9 to 12 deal with constraints. With the use of the relations 
$\downarrow, \uparrow, \not\downarrow, \not\uparrow$ introduced in Sect. 3.3
the rules are easy to formulate. For $e\con e'$ it is sufficient to
find two c-terms in the {\it SAS}'s verifying the relation $\downarrow$, what in
fact is equivalent to find a common totally defined c-term such that both
expressions $e$ and $e'$ can be 
reduced to it (observe the analogy with rule 5 of {\it CRWL}). For the
complementary constraint $\nocon$ we need to use all the information of {\it SAS}'s
in order to check the relation $\not\downarrow$ over all the possible pairs. The
explanation of rules 10 and 12 is quite similar.

Finally rules 13 and 14 provide together a formal definition of the function
${\it fails}$ supported by the notion of {\it SAS}. Notice that the {\it
  SAS}'s $\{\bot\}$ or $\{\bot,\fail\}$ do not provide enough information for
reducing a call to ${\it fails}$. The call ${\it fails}(e)$ is only reduced to
$\{{\it true}\}$ when every possible reduction of the expression $e$ is failed; and it
is reduced to $\{{\it false}\}$ there is some reduction of $e$ to some (possible
partial) c-term of the form $c(...)$ ($c\in DC$) or $X$.

\medskip

The next example shows a derivation of failure using the {\it CRWLF}-calculus.

\begin{example}
\label{ex:5}
\vspace*{0.2cm}

\noindent
Let us consider a program $\mathcal{P}$ with the constructors $z,s$ for natural
numbers, $[\ ]$ and `$:$' for lists (although we use Prolog-like
notation for them, that is, $[z,s(z) | L]$ represents the list $(z:(s(z):L))$) and also
the constructors $\mathsf{t,f}$ that represent the boolean values {\em true}
and {\em false}. Assume 
the functions $coin$ and $h$ defined in Sect. \ref{sec:cal_crwl} and
Sect. \ref{sec:examples} respectively and also the function $mb$ (member)
defined as:  

\begin{center}
$\begin{array}{l} 
  mb(X,[Y | \mathit{Ys}])\eqr \mathsf{t}\Leftarrow X\con Y\\
  mb(X,[Y | \mathit{Ys}])\eqr \mathsf{t}\Leftarrow mb(X,\mathit{Ys})\con \mathsf{t}
\end{array}$
\end{center}

If we try to evaluate the expression $mb(coin,[s(h)])$ it will
fail. Intuitively, from definition of $h$ the list in the second argument can be
reduced to lists of the form $[s(s(...))]$ and the possible values of $coin$,
$z$ and $s(z)$, do not belong to those lists. The {\it CRWLF}-calculus allows to build
a proof for this fact, that is, $mb(coin,[s(h)])\rec\{\fail\}$, in the following
way: by application of rule 4 the proof could proceed by generating {\it SAS}'s for
the arguments

\begin{center}
$coin\rec\{z,s(z)\}\quad (\varphi_1)\qquad\quad [s(h)]\rec\{ [ s(s(\bot))]\}\quad (\varphi_2)$
\end{center}

\noindent and then collecting the contributions of rules of $mb$ for each possible
combination of values for the arguments; for the pair $(z,[s(s(\bot))])$ the
contribution of the rules defining $mb$ (here we write $\rec_1$ to refer to the
first rule of {\em mb} and $\rec_2$ for the second) will be
\begin{center}
$mb(z,[s(s(\bot))])\rec_1\{\fail\}\quad (\varphi_3)\qquad\quad
mb(z,[s(s(\bot))])\rec_2\{\fail\}\quad (\varphi_4)$  
\end{center}

\noindent and for the pair $(s(z),[s(s(\bot)])$ we will have
\begin{center}
$mb(s(z),[s(s(\bot))])\rec_1\{\fail\}\quad (\varphi_5)\qquad\quad
mb(s(z),[s(s(\bot))])\rec_2\{\fail\}\quad (\varphi_6)$  
\end{center}

The full derivation takes the form:

 \begin{center}
$\mbox{\tiny{4}}\!\mifrac{
  \varphi_{1}\quad\varphi_{2}\quad\varphi_{3}\quad\varphi_{4}\quad
  \varphi_{5}\quad\varphi_{6}\quad}
{mb(coin,[s(h)])\re\{\fail\}}$
\end{center}

The {\it SAS} $\{\fail\}$ in the conclusion comes from the union of all the
contributing {\it SAS}'s of $\varphi_{_3}$, $\varphi_{_4}$, $\varphi_{_5}$ and
$\varphi_{_6}$. The statements $\varphi_{_1}$ to $\varphi_{_6}$ require of
course their own proof, which we describe now. At each step, we indicate by a
number on the left the rule of the calculus applied in each case: 

The derivation for $\varphi_{_1}$ is not difficult to build, and for
$\varphi_{_2}$ it is:
\begin{center}
$\fn$
\end{center}

For $\varphi_{_3}$ it can be done as follows:
\begin{center}
$\ft$
\end{center}

Here, the failure is due to a failure in the constraint $z\con s(s(\bot))$
of the used program rule, what requires to prove the complementary constraint $z\nocon
s(s(\bot))$ by rule (11). In this case there is a clear clash of constructors
($z$ and $s$).

For $\varphi_{_4}$ a derivation might be this one:
\begin{center}
$\faa$
\end{center}

The failure is due again to a failure in the constraint of the rule and in this
case the complementary constraint is $mb(z,[\ ])\nocon\ \mathsf{t}$. Now it is
involved the failure for the expression $mb(z,[\ ])$ that is proved by rule
(4) of the calculus. The {\it SAS}'s for the arguments only produce the
combination $(z,[\ ])$ and both rules of $mb$ fail over it by rule (8) of the
calculus.

The derivations for $\varphi_{_5}$ and $\varphi_{_6}$ are quite similar to 
those of $\varphi_{_3}$ and $\varphi_{_4}$ respectively. All the contributions
obtained from $\varphi_3,\varphi_4,\varphi_5$ and $\varphi_6$ are $\{\fail\}$,
and putting them together we obtain $\{\fail\}$ as an {\it SAS} for the original
expression $mb(coin,[s(h)])$, as it was expected.

\medskip

\end{example}

\section{Properties of {\it CRWLF}}
\label{sec:cons_monot}

In this section we explore some technical properties of the {\it CRWLF}-calculus
which are the key for proving the results of the next section, where we relate
the {\it CRWLF}-calculus to the {\it CRWL}-calculus.
In the following we assume a fixed program $\cal P$.

The non-determinism of the {\it CRWLF}-calculus allows to obtain different {\it SAS}'s for
the same expression. As an {\it SAS} for an expression is a finite approximation
to the denotation of the expression it is expected some kind of consistency
between {\it SAS}'s for the same expression. Given two of them, we cannot ensure that 
one {\it SAS} must be more defined than the other in the sense that all the
elements of the first are more defined than all of the second. For instance, two {\it
  SAS}'s for $coin$ are $\{\bot,s(z)\}$ and $\{ z,\bot\}$. 
 The kind of consistency for {\it SAS}'s that we can expect is the
following:

\begin{definition}[Consistent Sets of c-terms]
Two sets $\cb,\cb'\subseteq CTerm_{\bot,\fail}$ are {\em consistent} iff 
for all $t\in\cb$ there exists $t'\in\cb'$ (and vice versa, 
for all $t'\in\cb'$ there exists $t\in\cb$) such that $t$ and $t'$ are
consistent. 
\end{definition}

Our first result states that two different {\it SAS}'s for the same expression must
be consistent.

\begin{theorem}[Consistency of {\it SAS}]
\label{th:1}
  Given $e\in Term_{\bot,\fail}$, if $\mathcal{P}\conscrwlf e\rec\cb$ and
  $\mathcal{P}\conscrwlf e\rec\cb'$, then $\cb$ and $\cb'$ are consistent.
\end{theorem}

This result is a trivial corollary of part $a)$ of the following lemma.

\begin{lemma}[Consistency]
\label{lem:consist}
For any $e,e',e_1,e_2,e'_1,e'_2 \in Term_{\bot,\fail}$
  \begin{itemize}
  \item[$a)$] If $e, e'$ are consistent, 
  $\mathcal{P}\conscrwlf e\rec\cb$ and   $\mathcal{P}\conscrwlf  e'\rec\cb'$, 
then $\cb$ and $\cb'$ are consistent. 

  \item[$b)$] If $e_1,e'_1$ are consistent and $e_2,e'_2$ are also consistent, 
  then: $\mathcal{P}\conscrwlf e_1\diamondsuit e_2
        \Rightarrow \mathcal{P}\not\conscrwlf e'_1\tilde{\diamondsuit} e'_2$
  \end{itemize}
\end{lemma}

\begin{proof*}

For proving the consistency lemma we will split $b)$ into $b.1), b.2)$ and
also strengthen the lemma with a new part $c)$:

\begin{itemize}
\item[$b)$] If $e_1,e'_1$ are consistent and $e_2,e'_2$ are also consistent, then: 
  \begin{itemize}
  \item[$b.1)$] $\mathcal{P}\conscrwlf e_1\con e_2 \Rightarrow \mathcal{P}\not\conscrwlf
    e'_1\nocon e'_2$ 
  \item[$b.2)$] $\mathcal{P}\conscrwlf e_1\div e_2 \Rightarrow \mathcal{P}\not\conscrwlf
    e'_1\nodiv e'_2$ 
  \end{itemize}
  
\item[$c)$] Given $\overline{t},\overline{t}'\in CTerm_{\bot,\fail}\times ... \times
  CTerm_{\bot,\fail}$ pairwise consistent and $R\in\mathcal{P}_f$, if
  $\mathcal{P}\conscrwlf f(\overline{t})\recr\cb,f(\overline{t}')\recr\cb'$, then
  $\cb$ and $\cb'$ are consistent. 
\end{itemize}

Now we will prove $a),b)$ and $c)$ simultaneously  by induction on the size
  $l$ of the derivation for $e\rec\cb$ in $a)$, $e_1\con e_2$ in $b.1)$, $e_1\div
  e_2$ in $b.2)$ and $f(\overline{t})\recr\cb$ in $c)$. 

\noindent $\underline{l=1}$:
\begin{itemize}
\item[$a)$] The possible derivations in  one step are:
  \begin{itemize}
  \item[$\bullet$] $e\rec \{\bot\}$. This {\it SAS} is consistent with any other;
  \item[$\bullet$] $X\rec\{ X\}$. Then either  $e'=X$ or $e'=\bot$, so the possibilities for
    $\cb'$ are $\{ X\}$ or $\{\bot\}$, both consistent with $\{ X\}$;
  \item[$\bullet$] $c\rec\{ c\}$, where $c\in DC^0\cup\{\fail\}$. In this case $e'$ must be
    $c$ or $\bot$, whose possible {\it SAS}'s are $\{ c\}$ and $\{\bot\}$, that are
    consistent with $\{ c\}$.
  \end{itemize}

\item[$b$)] There is no derivation       of the form $e\con e'$ or $e\div e'$ in
   one step.
  
\item[$c)$] The possible derivations of the form $f(\overline{t})\recr\cb$ are:
  \begin{itemize}
  \item[$\bullet$] $f(\overline{t})\recr\{\bot\}$. This {\it SAS} is consistent with any other;
  \item[$\bullet$] $f(\overline{t})\recr\{\fail\}$, by means of rule 8, i.e., there exists some
    $R\equiv (f(\overline{s})\eqr e\Leftarrow \overline{C})\in \mathcal{P}_f$ and some $i$
    such that $s_i$ and $t_i$ have a $DC\cup\{\fail\}$-clash at some position
    $p$. The {\it SAS} $\cb'$ for $f(\overline{t}')$ using the function rule $R$ 
    must be done by one of the rules 5 to 8:
    
    \begin{itemize}
    \item if rule 5 is used then $\cb' = \{\bot\}$ that is consistent with $\cb$;
    \item rule 6 is not applicable: $t_i$ and $t'_i$ are consistent because 
      $\overline{t}$ and $\overline{t}'$ are pairwise consistent; then either $t'_i$ at
      position $p$ has the same constructor symbol as $t_i$ (and then the clash
      with $s_i$ remains), or $t'_i$ at $p$ or some of its ancestor positions
      has $\bot$. In both cases it is clear that there is not any c-instance of $R$
      for using rule 6;
    \item by rules 7 or 8 the {\it SAS} is $\{\fail\}$ that is consistent with the
      initial one $\{\fail\}$.      
    \end{itemize}
  \end{itemize}
\end{itemize}

\medskip

\noindent $\underline{l \Rightarrow l+1}$:

\begin{itemize}
\item[$a)$] In $l+1$ steps the possible derivations for $e\rec\cb$ are:
  \begin{itemize}
  \item[$\bullet$]
      $\mifrac{e_1\rec\cb_1\quad ... \quad e_n\rec\cb_n}
      {e=c(e_1,...,e_n)\rec\{ c(t_1,...,t_n) | \overline{t}\in\cb_1\times
      ...\times\cb_n\}} $
    by rule 3, where $c\in DC^n$ ($n>0$). Then either  $e'=\bot$, whose only
    possible {\it SAS} is $\{\bot\}$, that is consistent with any other, or
    $e'=c(e'_1,...,e'_n)$ with $e_i$ and $e'_n$ being consistent 
    for $i\in \{1,\ldots,n\}$ and the {\it SAS} is produced by rule 3:
     \[ \mifrac{e'_1\rec\cb'_1\quad ... \quad e'_n\rec\cb'_n}
      {e'=c(e'_1,...,e'_n)\rec\{ c(t'_1,...,t'_n) | \overline{t}'\in\cb'_1\times
      ...\times\cb'_n\}}\]
   By i.h. $\cb'_i$ is consistent with $\cb_i$ for all
      $i\in\{1,...,n\}$ and then it is clear that $\cb$ and $\cb'$ are also
      consistent. 
    
  \item[$\bullet$]
      $\mifrac{e_1\rec\cb_1\quad ... \quad e_n\rec\cb_n\quad f(\overline{t})\recr\cb_{R,\overline{t}}}
      {e=f(e_1,...,e_n)\rec\bigcup_{R\in\mathcal{P}_f,\overline{t}\in\cb_1\times
        ... \times\cb_n} \cb_{R,\overline{t}}}$
    by rule 4. Then either  $e'=\bot$ whose only possible {\it SAS} is $\{\bot\}$
    that is consistent with any other, or $e'=f(e'_1,...,e'_n)$ with $e_i,
    e'_i$ consistent for all $i\in\{ 1,...,n\}$. If the {\it SAS} for $e'$ is
    generated by rule 1 of the calculus, the result would be clear and for rule
    4 we have
     $\mifrac{e'_1\rec\cb'_1\quad ... \quad e'_n\rec\cb'_n\quad f(\overline{t}')\recr\cb_{R,\overline{t}'}}
      {e'=f(e'_1,...,e'_n)\rec\bigcup_{R\in\mathcal{P}_f,\overline{t}'\in\cb'_1\times
        ... \times\cb'_n} \cb_{R,\overline{t}'}}$

    By i.h. $\cb_i$ and $\cb'_i$ are consistent for all $i\in\{1,...,n\}$, what
    means that for each $\overline{t}\in\cb_1\times ... \times\cb_n$ there exists
    $\overline{t}'\in\cb'_1\times ... \times\cb'_n$ consistent with
    $\overline{t}$. Again by i.h. we have that each {\it SAS}
    $\cb_{R,\overline{t}}$ is consistent with $\cb_{R,\overline{t}'}$ and it can
    be easily proved that
    $\cb =\bigcup_{R\in\mathcal{P}_f,\overline{t}\in\cb_1\times
      ... \times\cb_n}\cb_{R,\overline{t}}$ is then consistent with
    $\cb' =\bigcup_{R\in\mathcal{P}_f,\overline{t}'\in\cb'_1\times
      ... \times\cb'_n}\cb_{R,\overline{t}'}$.

  \item[$\bullet$]
      $\mifrac{e_1\rec\{\fail\}}
      {e={\it fails}(e_1)\rec\{{\it true}\}}$
    by rule 13. If $e'=\bot$ the result is clear, else
    $e'={\it fails}(e'_1)$. Then the {\it SAS} for $e'$ requires to obtain an {\it SAS}
    $\cb'$ for $e'_1$. By i.h., $\cb'$ must be consistent with $\{\fail\}$ what
    means that $\cb'=\{\fail\}$ or $\cb'=\{\bot\}$. Then, the possible {\it
    SAS}'s for $e'$ are $\{\bot\}$ and $\{{\it true}\}$ (by the same rule 13), both
    consistent with $\{{\it true}\}$.

  \item[$\bullet$]
      $\mifrac{e_1\rec\cb_1}
      {e={\it fails}(e_1)\rec\{{\it false}\}}$
    by rule 14, such that there exists $t\in\cb_1$ with $t\not =\bot$, $t\not =\fail$. If $e'=\bot$
    the result is clear. Otherwise, if $e'={\it fails}(e'_1)$ the {\it SAS} for $e'$
    must be obtained by one of the rules 1, 13 or 14. By rule 1 it would be
    $\{\bot\}$ consistent with any other one; rule 13 would need to obtain the {\it SAS}
    $\{\fail\}$ for $e'_1$, but this is not possible because it must be
    consistent with $\cb_1$ by i.h., so rule 13 is not applicable; and rule 14
    would provide the {\it SAS} $\{{\it false}\}$ for $e'$, consistent with itself.

  \end{itemize}

\item[$b.1)$] If we have a derivation for $e_1\con e_2$ by rule 9, there exist two {\it SAS}'s
  $\cb_{e_1}$ and  $\cb_{e_2}$ such that $e\rec \cb_{e_1}, e_2\rec\cb_{e_2}$ and there exist
  $t\in \cb_{e_1}, t'\in\cb_{e_2}$ with $t\downarrow t'$.

  Now, let $e'_1, e'_2$ be consistent with $e_1,e_2$ respectively, and assume
  that $e'_1\nocon e'_2$ can be proved. We reason by
  contradiction. Since $e'_1\nocon e'_2$ is provable, we can prove $e'_1\rec\cb_{e'_1},
  e'_2\rec\cb_{e'_2}$ such that for all $s\in\cb_{e'_1}, s'\in\cb_{e'_2}$ it will be
  $s\not\downarrow s'$. 

  By i.h. $\cb_{e_1}$ is consistent with $\cb_{e'_1}$, what implies that there exists
  $u\in\cb_{e'_1}$ consistent with $t$, and then there exists $v$ such that
  $v\sqsupseteq u, v\sqsupseteq t$. In a similar way, there exists $u'\in\cb_{e'_2}$
  consistent with $t'$, so there exists $v'$ such that $v'\sqsupseteq u',
  v'\sqsupseteq t'$.

  As $u\in\cb_{e'_1}$ and $u'\in\cb_{e'_2}$ we would have $u\not\downarrow u'$; by
  monotonicity of $\not\downarrow$ we have $v\not\downarrow v'$, what implies 
  $\neg(v\downarrow v')$. But monotonicity of $\downarrow$, together with
  $t\downarrow t', v\sqsupset t, v'\sqsupset t'$, implies $v\downarrow v'$, what
  is a contradiction.

\item[$b.2)$] The case of $e_1\div e_2$ proceeds similarly to $b.1)$, using in
  this case monotonicity of $\uparrow$ and $\not\uparrow$.

  \end{itemize}

\item[$c)$] In $l+1$ steps the possible derivations for $f(\overline{t})\recr\cb$
  where $R\equiv (f(\overline{s})\eqr e\Leftarrow\overline{C})$, are:
  \begin{itemize}
  \item
    $\mifrac{e\theta\rec\cb\quad \overline{C}\theta}
    {f(\overline{t})\recr\cb}$
  by rule 6, using the c-instance $R\theta$ ($\theta\in
  CSubst_{\bot,\fail}$), such that $\overline{t}=\overline{s}\theta$. The derivation
  $f(\overline{t}')\recr\cb'$ must be done by
  one of the rules 5 to 8:
  \begin{itemize}
  \item if $f(\overline{t}')\recr\{\bot\}$ by rule 5, it is clear that this {\it SAS}
    is consistent with $\cb$;
  \item if the derivation is done by rule 6, it will have the form
      $\mifrac{e\theta'\rec\cb'\quad \overline{C}\theta'}
      {f(\overline{t}')\recr\cb'}$
    using a c-instance $R\theta'$ of $R$. In particular, we have
    $\overline{t}'=\overline{s}\theta'$ and we also had 
    $\overline{t}=\overline{s}\theta$. As $\overline{t}$ and $\overline{t}'$
    are pairwise consistent, and $var(e)\subseteq var(\overline{s})$ it is not
    difficult to see that $e\theta$ and $e\theta'$ must be consistent. Then
    by i.h. (part $a)$) we deduce that $\cb$ and $\cb'$ are consistent {\it
    SAS}'s.

  \item rule 7 is not applicable: suppose that we have the derivation 
     $\mifrac{\widetilde{C_i}\theta'}
      {f(\overline{t}')\recr\{\fail\}}$
    using a c-instance $R\theta'$ of $R$ and $C_i\theta'$ being a constraint of
    $\overline{C}\theta$. Analogously to the previous case,
    we have that both members of $C_i\theta'$ are consistent with the corresponding ones of
    $C_i\theta$; as $C_i\theta$ is provable then by i.h. (part $b)$),
    $\widetilde{C}_i\theta'$ is not provable, what means that rule 7 cannot be applied.

  \item rule 8 is not applicable: there cannot be a pair of $\overline{t}'$ and
    $\overline{s}$ with a $DC\cup\{\fail\}$-clash because then the corresponding
    pair of $\overline{t}'$ and $\overline{s}\theta =\overline{t}$ would have
    the same clash (the substitution $\theta$ cannot make disappear the clash).

  \end{itemize}

\item 
    $\mifrac{\widetilde{C_i}\theta}
    {f(\overline{t})\recr\{\fail\}}$
  by rule 7, being $R\theta$ a c-instance of the rule $R$ such that
  $\overline{t}=\overline{s}\theta$. The derivation $f(\overline{t}')\recr\cb'$ can
  be done by one of the rules 5 to 8:
  \begin{itemize}
  \item by rule 5, the {\it SAS} is $\{\bot\}$ that is consistent with any other;
  \item it is not possible to use rule 6 because we would need to prove a
    constraint $C_i\theta'$ of a c-instance $R\theta'$ of $R$. As
    $\overline{s}\theta = \overline{t}$ and $\overline{s}\theta' =
    \overline{t}'$ are pairwise consistent and $var(C_i)\subseteq
    var(\overline{s})$, both members of $C_i\theta$ and $C_i\theta'$ will be
    also consistent. Then by i.h. (part $b)$), as $\widetilde{C_i}\theta$ is provable,
    $C_i\theta'$ will not be provable.
  \item if 7 or 8 applies we will have $\cb'=\{\fail\}$ that is
    consistent with $\cb =\{\fail\}$ (in fact, 8 would not be applicable). 
$\mathproofbox$
  \end{itemize}
\end{itemize}
\end{proof*}

\medskip

As a trivial consequence of part $b)$ we have:

\begin{corollary}
 $\mathcal{P}\conscrwlf e\diamondsuit e' \Rightarrow 
  \mathcal{P}\not\conscrwlf e\tilde{\diamondsuit} e'$, for all 
    $e,e'\in Term_{\bot,\fail}$
\end{corollary}

This justifies indeed our description of $\nocon$ and $\nodiv$ as 
computable approximations to the negations of $\con$ and $\div$. 

\medskip

Another desirable property of our calculus is {\em monotonicity}, that we can
informally understand in this way: the information that can be extracted from an expression
cannot decrease when we add information to the expression itself. This
applies also to the case of constraints: if we can prove a constraint and we consider more
defined terms in both sides of it, the resulting constraint must be also
provable. Formally:

\begin{proposition}[Monotonicity of {\it CRWLF}]
\label{prop:2}
For $e,e',e_1,e_2,e'_1,e'_2 \in Term_{\bot,\fail}$
  \begin{itemize}
  \item[$a)$] If $e\sqsubseteq e'$ and $\mathcal{P}\conscrwlf e\rec\cb$, then 
$\mathcal{P}\conscrwlf e'\rec\cb$
  \item[$b)$] If $e_1\sqsubseteq e'_1$, $e_2\sqsubseteq e'_2$ and
 $\mathcal{P}\conscrwlf e_1\diamondsuit e_2$
    then $\mathcal{P}\conscrwlf e'_1\diamondsuit e'_2$, 
    where \mbox{$\diamondsuit\in\{\con,\nocon,\div,\nodiv\}$}
  \end{itemize}
\end{proposition}

\begin{proof*} Again we need to strengthen the result with a new part $c)$
  \begin{itemize}
  \item[$c)$] Given $\overline{t},\overline{t}'\in CTerm_{\bot,\fail}\times ...\times
    CTerm_{\bot,\fail}$ such that $t_i\sqsubseteq t'_i$ for all $i\in\{ 1,...,n\}$ and
    $R\in\mathcal{P}_f$, if $f(\overline{t})\recr\cb$ then
    $f(\overline{t}')\recr\cb$
  \end{itemize}

We will prove parts $a),b)$ and $c)$ simultaneously by induction  on the size 
$l$ of the derivation for $e\rec\cb$ in $a)$, $e_1\diamondsuit e_2$ in $b)$ and
$f(\overline{t})\recr\cb$ in $c)$:

\noindent $\underline{l=1}$:
\begin{itemize}
\item[$a)$] The derivation of $e\rec\cb$ in  one step can be:
  \begin{itemize}
  \item[$\bullet$] $e\rec\{\bot\}$, and it is clear that also $e'\rec \{\bot\}$
  \item[$\bullet$] $X\rec \{ X\}$: then $e=e'=X$
  \item[$\bullet$] $c\rec\{ c\}$, $c\in DC^0$: then $e=e'=c$
  \end{itemize}

\item[$b)$] For $e_1\diamondsuit e_2$ there are not possible derivations in  one steps.

\item[$c)$] For $f(\overline{t})\recr\cb$ the derivations can be:
  \begin{itemize}
  \item[$\bullet$]  $f(\overline{t})\recr \{\bot\}$, using rule 5. Then for all $\overline{t}'$ we have
    $f(\overline{t}')\recr\{\bot\}$
  \item[$\bullet$]  $f(\overline{t})\recr \{\fail\}$, using rule 8. Then $R\equiv (f(\overline{s})\eqr e
    \Leftarrow \overline{C})$ and $\overline{t}$ and $\overline{s}$ have a
    $DC$-clash at some position. If $\overline{t}\sqsubseteq\overline{t}'$ then 
    $\overline{t}'$ and $\overline{s}$ have the same clash, and rule 8 allows to
    prove also $f(t')\recr\{\fail\}$.
  \end{itemize}
\end{itemize}

\noindent $\underline{l \Rightarrow l+1}$:

\begin{itemize}
\item[$a)$] We distinguish three cases for the derivation of $e\rec\cb$:
  \begin{itemize}
  \item[$\bullet$]  $e= c(e_1,...,e_n)$. Then the derivation of $e\rec\cb$ must use the rule 3
    and take the form:
      $\mifrac{e_1\rec\cb_1\ ...\ e_n\rec\cb_n}
      {c(e_1,...,e_n)\rec\{ c(t_1,...,t_n) | \overline{t}\in\cb_1\times ...\times\cb_n\}}$
    Since $e\sqsubseteq e'$, $e'$ must take the form  $e'=c(e'_1,...,e'_n)$ with
    $e_1\sqsubseteq e'_1,...,e_n\sqsubseteq e'_n$. By i.h. we have
    $e'_1\rec\cb_1,...,e'_n\rec\cb_n$ and with the same rule 3 
    we can build a derivation for $c(\overline{e}')\rec\cb$.

  \item[$\bullet$]  $e=f(e_1,...,e_n)$. Then the derivation of $e\rec\cb$ must use rule 4 and
    take the form:
      $\mifrac{e_1\rec\cb_1\ ...\ e_n\rec\cb_n\ f(\overline{t})\recr\cb_{R,\overline{t}}}
      {f(e_1,...,e_n)\rec\bigcup_{R\in \mathcal{P}_f,\overline{t}\in\cb_1\times
        ...\times\cb_n} \cb_{R,\overline{t}}}$
     $e'$ must take the form $e'=f(e'_1,...,e'_n)$ with $e_i\sqsubseteq
    e'_i$. By i.h. we will have $e'_1\rec\cb_1,...,e'_n\rec\cb_n$ and then we have
    the same tuples $\overline{t}$, the same {\it SAS}'s $\cb_{R,\overline{t}}$
    and finally the same {\it SAS} for $f(\overline{e}')$.

  \item[$\bullet$]  $e={\it fails}(e_1)$. Then $e'={\it fails}(e'_1)$. The derivation $e\rec\cb$ must
    be done by one of the rules 13 or 14, that require to obtain an {\it SAS} for
    $e_1$. By i.h. if $e_1\rec\cb_1$ then $e'_1\rec\cb_1$ and then the same rule
    (and only that) is applicable to obtain the same {\it SAS} for $e'$, that
    will be $\{{\it true}\}$ if rule 13 is applicable or $\{{\it false}\}$ if rule 14  is applied.
  \end{itemize}

\item[$b)$] The derivation $e_1\diamondsuit e_2$ with
  $\diamondsuit\in\{\con,\nocon,\div,\nodiv\}$ will be done by generating the 
  {\it SAS}'s $e_1\rec\cb_1$ and  $e_2\rec\cb_2$. By i.h. we have $e'_1\rec\cb_1,
  e'_2\rec\cb_2$ and then it is clear that $e'_1\diamondsuit e'_2$ is also provable.

\item[$c)$] We distinguish the following cases according to the rule used for the derivation of
  $f(\overline{t})\recr\cb$:
  \begin{itemize}
  \item[$\bullet$] By rule 6 the derivation would be:
    $\mifrac{e\theta\rec\cb\ \overline{C}\theta}
    {f(t_1,...,t_n)\recr\cb}$
 where the rule $R$ is $R\equiv (f(s_1,...,s_n)\eqr e\Leftarrow \overline{C})$ and $\theta\in
  Subst_{\bot,\fail}$ such that $\overline{s}\theta =\overline{t}$.

  We will show that the same rule 6 is applicable for generating an {\it SAS} for
  $f(\overline{t}')$ being $t_i\sqsubseteq t'_i$ for all $i\in\{1,...,n\}$. The
  idea is that if $t_i\sqsubseteq t'_i$ then $t'_i$ is the result of replacing
  some subterms $\bot$ of $t_i$ by c-terms more defined than $\bot$. As $s_i\in
  CTerm$ then the corresponding positions or some ancestors must have variables
  in $s_i$. Then we can get a substitution $\theta'\in CSubst_{\bot,\fail}$ such
  that $\theta\sqsubseteq\theta'$ and $s_i\theta' = t'_i$. A formal
  justification of this fact may be done by 
  induction on the syntactic structure of $t_i$ and, as $\overline{s}$ is a
  linear tuple, the result can be extended in such a way that
  $\overline{s_i}\theta'=\overline{t}'$. 

  We also have that $e\theta \sqsubseteq e\theta'$, so by i.h. we have
  $e\theta'\rec\cb$. As the constraints $\overline{C}\theta$ are provable and
  $\theta\sqsubset\theta'$, then
  by i.h. $b)$, the constraints $\overline{C}\theta'$ will also be provable. So
  we can build a derivation for $f(\overline{t}')\recr\cb$ by rule 6.

  \item[$\bullet$] By rule 7 the derivation would be:\ \
   $\mifrac{\widetilde{C}_i\theta}
    {f(\overline{t})\recr\{\fail\}}$
 where the rule $R$ is $R\equiv (f(\overline{s})\eqr e\Leftarrow  C_1,...,C_n),
  i\in\{1,...,n\}$ and $\theta\in CSubst_{\bot,\fail}$ is such that
  $\overline{s}\theta = \overline{t}$. 

  As $\overline{t}\sqsubseteq \overline{t}'$, in a similar way as before there
  exists $\theta'$ such that $\overline{s}\theta' =\overline{t}'$
  and by i.h. we can prove $\widetilde{C}_i\theta'$, what implies that we can build
  the  derivation for $f(\overline{t}')\recr \{\fail \}$, using rule 7.
$\mathproofbox$
  \end{itemize}
\end{itemize}
\end{proof*}

\medskip

\noindent {\bf Remark:} Monotonicity, as stated in Prop. \ref{prop:2}, refers to
the degree of evaluation of expressions and does not contradict
the well known fact that negation as failure is a non-monotonic reasoning rule. In our setting it 
is also clearly true that, if we `define more' the functions (i.e, we refine the program,
by adding new rules to it), an expression can become reducible when
it was previously failed.

\medskip

The next property says that what is true for free variables is also true
for any possible (totally defined) value, i.e., provability in {\it CRWLF} is
closed under total substitutions.

\begin{proposition}
\label{prop:3}
  For any $\theta\in CSubst$, $e,e'\in Term_{\bot,\fail}$  
  \begin{itemize}
  \item[$a)$] $\mathcal{P}\conscrwlf e\rec\cb \Rightarrow
              \mathcal{P}\conscrwlf e\theta\rec\cb\theta$
  \item[$b)$] $\mathcal{P}\conscrwlf e\diamondsuit e'\Rightarrow
             \mathcal{P}\conscrwlf e\theta\diamondsuit e'\theta$
  \end{itemize}

\end{proposition}

\begin{proof*}

Again we need to strengthen the result, with a new part $c)$:

\begin{itemize}
\item[c)] $f(\overline{t})\recr\cb\Rightarrow
  f(\overline{t})\theta\recr\cb\theta$, for any $\overline{t}\in
  CTerm_{\bot,\fail}\times ...\times 
  CTerm_{\bot,\fail}$
\end{itemize}

We prove simultaneously the three parts by induction on the size $l$ of the
derivations. 

\noindent $\underline{l=1}$: in  one step we can have the derivations $e\rec\{\bot\},c\rec\{ c\}$
($c\in DC^0 \cup \{\fail\}$) and $X\rec\{ X\}$. The property is obvious for the first two and the
third follows from the fact that if $t\in CTerm_{\bot,\fail}$ then $t\rec\{ t\}$
is provable (this can be proved by induction on the depth of the term
$t$). Notice that $X\theta\in CTerm\subset CTerm_{\bot,\fail}$, so
  $X\theta\rec\{ X\theta\}$.

\noindent $\underline{l \Rightarrow l+1}$: now we can have the following derivations:
\begin{itemize}
\item by rule 3 we have
    $\mifrac{e_1\rec \cb_1\quad ...\quad e_n\rec \cb_n}
    {c(e_1,...,e_n)\rec \{c(t_1 ,...,t_n)\ | \ \overline{t}\in \cb_1\times
    ...\times\cb_n \}}$
 By i.h. we have $e_i\theta\rec\cb_i\theta$ for all $i\in\{1,...,n\}$ and again by rule
  3 we can build a derivation for $c(\overline{e})\theta\rec\{c(t_1,...,t_n)\theta\ |\
  \overline{t}\theta\in \cb_1\theta\times ...\times\cb_n\theta \}$

\item by rule 4 we have
    $\mifrac{e_1\rec \cb_1\quad ...\quad e_n\rec \cb_n\quad ...\quad
    f(\overline{t})\recr\cb_{R,\overline{t}}\quad ...}
    {f(e_1,...,e_n)\rec \bigcup_{R\in\mathcal{P}_f ,\overline{t}\in\cb_1\times
      ...\times\cb_n} \cb_{R,\overline{t}}}$
  By i.h. we have $e_i\theta\rec\cb_i\theta$ for all $i\in\{1,...,n\}$ and
  $f(\overline{t})\theta\recr\cb_{R,\overline{t}}\theta$ for each
  $\overline{t}\theta\in \cb_1\theta\times ...\times\cb_n\theta \}$ and each
  rule $R\in\mathcal{P}_f$. So we can get a derivation for
  $f(\overline{e})\theta\rec \bigcup_{R\in\mathcal{P}_f,\overline{t}\theta\in \cb_1\theta\times
    ...\times\cb_n\theta} \cb_{R,\overline{t}\theta}$

\item by rule 5 we have
    $\mifrac{}{f(\overline{t})\recr\{ \bot\}}$
  and it is clear $f(\overline{t})\theta\recr\{ \bot\}$

\item by rule 6 we have       
   $\mifrac{e\theta'\rec\cb\quad \overline{C}\theta'}
    {f(\overline{t})\recr\cb}$
  where $(f(\overline{s})\eqr e \Leftarrow \overline{C})\in R$ and $\theta'\in
  CSubst_{\bot,\fail}$ is such that $f(\overline{t})=f(\overline{s})\theta'$. For
  the call $f(\overline{t})\theta$ we can get the appropriate c-instance by
  composing $\theta'$ and $\theta$, so $f(\overline{t})\theta =
  f(\overline{s})\theta'\theta$. By i.h. we have $e\theta'\theta\rec\cb\theta$
  and $\overline{C}\theta'\theta$, and then 
  $f(\overline{t})\theta\recr\cb\theta$ by the same rule 6.

\item by rule 7 we have
    $\mifrac{e_i\theta'\tilde{\diamondsuit} e'_i\theta'}
    {f(\overline{t})\recr \{ F\}}$
  where $(f(\overline{s})\eqr e \Leftarrow \overline{C})\in R$ and $\theta'\in
  CSubst_{\bot,\fail}$ is such that $f(\overline{t})=f(\overline{s})\theta'$. As
  before, for the call $f(\overline{t})\theta$ we can get the appropriate
  c-instance by composing $\theta'$ and $\theta$, so $f(\overline{t})\theta = 
  f(\overline{s})\theta'\theta$. By i.h. we have
  $e_i\theta'\theta\tilde{\diamondsuit} e'_i\theta'\theta$, and then 
  $f(\overline{t})\theta\recr \{ F\}$

\item by rule 8 we have 
    $\mifrac{}{f(t_1,...,t_n)\recr \{ F\}}$
  where $R\equiv (f(s_1,...,s_n)\eqr  e \Leftarrow\overline{C})$ and such that
  $t_i$ and $s_i$ have a $DC\cup\{\fail\}$-clash for some
  $i\in\{1,...,n\}$. It is clear that $t_i\theta$ and $s_i$ will have the same
  clash so $f(\overline{t})\theta\rec_r\{\fail\}$

\item by rules 9 to 12, the derivation would have the form $e\diamondsuit e'$. By i.h. we
  have $e\theta\rec\cb\theta\quad 
  e'\theta\rec\cb'\theta$. Now, if we take $t\in\cb, t'\in\cb'$ and $t\Re t'$
  holds (where $\Re\in\{\downarrow,\uparrow,\not\downarrow,\not\uparrow\}$),
  then $t\theta\Re t'\theta$ also holds, by Prop. \ref{prop:1}. It follows that
  $e\theta\diamondsuit e'\theta$.

\item by rule 13 (or rule 14), it must be $e={\it fails}(e_1)$. This rule requires to
  obtain an {\it SAS} for $e_1$, say $e_1\rec\cb_1$. Then by
  i.h. $e_1\theta\rec\cb_1\theta$ and it is clear that rule 13 
  will be applicable to derive $e\rec\{{\it true}\}$ (or $e\rec\{{\it false}\}$ by rule 14).
$\mathproofbox$
\end{itemize}
\end{proof*}

\section{{\it CRWLF} related to {\it CRWL}}
\label{subsec:rel}

The {\it CRWLF}-calculus has been built as an {\em extension} of {\it CRWL} for dealing with
{\em failure}. Here we show that our aims have been achieved with respect to 
these two emphasized aspects. In order to establish the relations between both
calculus we consider in this section the class of programs defined for {\it
  CRWL}, i.e., rules cannot use the function ${\it fails}$. This means that
rules 13 and 14 of the {\it CRWLF}-calculus are not considered here.

First, we show that the {\it CRWLF}-calculus indeed extends {\it CRWL}.
Parts {\em a)} and {\em b)} of the next result show that statements $e \rec \cb$ generalize
approximation statements $e \to t$ of {\it CRWL}. Parts  {\em c)} and {\em d)}
show that {\it CRWLF} and {\it CRWL} are able to prove exactly the same 
joinabilities and divergences (if $\fail$ is ignored for the comparison).

\begin{proposition}
\label{prop:4}
For any $e,e'\in Term_{\bot,\fail}$
  \begin{itemize}
  \item[$a)$] $\mathcal{P}\conscrwlf e\rec\cb \Rightarrow \forall t\in\cb, \mathcal{P}\conscrwl
    \hat{e}\to\hat{t}$
  \item[$b)$] $\mathcal{P}\conscrwl \hat{e}\to t\Rightarrow \exists\cb$ such that
    $t\in\cb$ and $\mathcal{P}\conscrwlf e\rec\cb$ 
  \item[$c)$] $\mathcal{P}\conscrwlf e\con e'\Leftrightarrow\mathcal{P}\conscrwl
    \hat{e}\con\hat{e}'$
  \item[$d)$] $\mathcal{P}\conscrwlf e\div e'\Leftrightarrow\mathcal{P}\conscrwl
    \hat{e}\div\hat{e}'$ 
  \end{itemize}  
\end{proposition}

In order to prove the property we split it into two separate lemmas.
 The first one contains $a)$, the right implication of $c)$ and $d)$ and a new
 part $e)$: 

\begin{lemma} Let $\mathcal{P}$ a {\it CRWLF}-program. Then:
  \begin{itemize}
  \item[$a)$] $\mathcal{P}\conscrwlf e\rec\cb \Rightarrow \forall t\in\cb, \mathcal{P}\conscrwl
    \hat{e}\to\hat{t}$
  \item[$c)$] $\mathcal{P}\conscrwlf e\con e'\Rightarrow\mathcal{P}\conscrwl
    \hat{e}\con\hat{e}'$
  \item[$d)$] $\mathcal{P}\conscrwlf e\div e'\Rightarrow\mathcal{P}\conscrwl
    \hat{e}\div\hat{e}'$ 
  \item[$e)$] Given $\overline{t}\in CTerm_{\bot,\fail}\times ...\times
  CTerm_{\bot,\fail}$ and $R\in \mathcal{P}_f$: $\mathcal{P}\conscrwlf
  f(\overline{t})\recr\cb\Rightarrow \forall t\in\cb, \mathcal{P}\conscrwl
  \widehat{f(\overline{t})}\to \hat{t}$  

  \end{itemize}  
\end{lemma}

\begin{proof*}
We prove simultaneously all the parts by induction on the size $l$ of
the corresponding derivation:

\noindent $\underline{l=1}$: the derivation can be:
\begin{itemize}
\item $\mathcal{P}\conscrwlf e\rec\{\bot\}$, and we have $\mathcal{P}\conscrwl
  \hat{e}\to\bot$
\item $\mathcal{P}\conscrwlf X\rec\{ X\}$, we have $\hat{X}=X$ and
  $\mathcal{P}\conscrwl X\to X$
\item $\mathcal{P}\conscrwlf c\rec \{ c\}$, where $c\in DC^0$ and we have
  $\hat{c}=c$ and $\mathcal{P}\conscrwl c\to c$
\item $\mathcal{P}\conscrwlf \fail\rec\{ \fail\}$, we have $\hat{\fail}=\bot$ and
  $\mathcal{P}\conscrwl \bot\to\bot$
\item $\mathcal{P}\conscrwlf f(\overline{t})\recr\{ \bot\}$, and we have
  $\mathcal{P}\conscrwl \widehat{f(\overline{t})}\to \bot$ 
\item $\mathcal{P}\conscrwlf f(\overline{t})\recr\{ \fail\}$, and we have
  $\mathcal{P}\conscrwl \widehat{f(\overline{t})}\to \bot$
\end{itemize}

\medskip

\noindent $\underline{l \Rightarrow l+1}$: the derivation can be:

\begin{itemize}
\item $\mathcal{P}\conscrwlf c(e_1,...,e_n)\rec\{...,c(t_1,...,t_n),...\}$,
  then by the rule 3 of {\it CRWLF}, it must be $\mathcal{P}\conscrwlf e_i\rec 
  \{...,t_i,...\}$. By i.h. we have  $\mathcal{P}\conscrwl \hat{e}_i\to \hat{t}_i$ and
  then we can build the derivation 
  $\mathcal{P}\conscrwl \widehat{c(e_1,...,e_n)}\to \widehat{c(t_1,...,t_n)}$,
  by the rule 3 of {\it CRWL}.

\item $\mathcal{P}\conscrwlf f(e_1,...,e_n)\rec\{ ...,t,...\}$. This derivation must
  use the rule 4 of {\it CRWLF}, and then we will have the derivations
  $\mathcal{P}\conscrwlf e_i\rec\mathcal{C}_i$ for all $i\in\{1,...,n\}$ and
  $f(\overline{t})\recr \cb_{R,\overline{t}}$. It must be $t\in
  \cb_{R,\overline{t}}$ for some $\overline{t}\in \cb_1\times 
  ...\times\cb_n$ and $R\in \mathcal{P}_f$. By i.h. we will have
  $\mathcal{P}\conscrwl \widehat{f(\overline{t})}\to \hat{t}$

\item $\mathcal{P}\conscrwlf f(\overline{t})\recr\cb$ by rule 6 of
  {\it CRWLF}, for which we take $R\equiv (f(\overline{s})\eqr e\Leftarrow \overline{C}),
  \theta\in CSubst_{\bot,\fail}$ such that $\overline{s}\theta 
  =\overline{t}$. We can define $\theta'\in CSubst_{\bot}$ as $X\theta'
  =\bot$ if $X\theta =\fail$ and $X\theta' = X\theta$, in other case. So we have
  $\overline{s}\theta' =\hat{\overline{t}}$, $e\theta'=\widehat{e\theta}$ and
  $\overline{C}\theta' = \widehat{\overline{C}\theta}$. Now we can take
  $(f(\overline{s})\eqr e\Leftarrow \overline{C})\theta'\in [ \mathcal{P}]_{\bot}$. We also have
  $e\theta\rec\cb$ and if $t\in\cb$ by i.h. we have $\widehat{e\theta}\to
  \hat{t}$, or what is the same, $e\theta'\to \hat{t}$. Also by
  i.h. $\overline{C}\theta'=\widehat{\overline{C}\theta}$ is provable within {\it CRWL}, and
  therefore $\hat{f(\overline{t})}\to\hat{t}$ by rule 4 of {\it CRWL}.

\item $\mathcal{P}\conscrwlf f(\overline{t})\recr \{\fail\}$ and we have
  $\mathcal{P}\conscrwl \widehat{f(\overline{t})}\to \bot$

\item $\mathcal{P}\conscrwlf e\con e'$ using the rule 9 of {\it CRWLF}. Then we will have
   $\mathcal{P}\conscrwlf e\rec\cb$, $\mathcal{P}\conscrwlf
  e'\rec\cb'$ and there exist $t\in\cb, t'\in\cb'$ such that $t\downarrow
  t'$ (by definition of $\downarrow$ it is easy to see that $t=t'$). By
  i.h. we have $\mathcal{P}\conscrwl \hat{e}\to \hat{t}$ and $\mathcal{P}\conscrwl
  \hat{e}'\to \hat{t}$ and by rule 5 of {\it CRWL} we have
  $\mathcal{P}\conscrwl \hat{e}\con \hat{t}$. 

\item $\mathcal{P}\conscrwlf e\div e'$ using the rule 10 of {\it CRWLF}. Then we will have
   $\mathcal{P}\conscrwlf e\rec\cb$, $\mathcal{P}\conscrwlf
  e'\rec\cb'$ and there exist $t\in\cb, t'\in\cb'$ such that $t\uparrow
  t'$. By definition of $\uparrow$, $t$ and $t'$ have a $DC$-clash. By i.h. we
  have $\mathcal{P}\conscrwl \hat{e}\to \hat{t}$ and 
  $\mathcal{P}\conscrwl \hat{e}'\to \hat{t}$ and by rule 5 of {\it CRWL} we have
  $\mathcal{P}\conscrwl \hat{e}\div \hat{t}$. 
$\mathproofbox$
\end{itemize}
\end{proof*}

\medskip

We now state the second lemma for Proposition \ref{prop:4}, in which the part $b)$ and
the left implications of $c)$ and $d)$ will be proved.

\begin{lemma}
For any $e,e'\in Term_{\bot,\fail}$
  \begin{itemize}
  \item[$b)$] $\mathcal{P}\conscrwl \hat{e}\to t\Rightarrow \exists\cb$ such that
    $t\in\cb$ and $\mathcal{P}\conscrwlf e\rec\cb$ 
  \item[$c)$] $\mathcal{P}\conscrwl
    \hat{e}\con\hat{e}'\Rightarrow\mathcal{P}\conscrwlf e\con e'$
  \item[$d)$] $\mathcal{P}\conscrwl\hat{e}\div\hat{e}'\Rightarrow
    \mathcal{P}\conscrwlf e\div e'$ 
  \end{itemize}  
\end{lemma}

\begin{proof*}
We prove the three parts simultaneously by induction on the size $l$
of the derivation: 

\noindent $\underline{l=1}$: the derivation can be:
\begin{itemize}
\item $\mathcal{P}\conscrwl \hat{e}\to\bot$ and it is clear that $\mathcal{P}\conscrwlf
  e\rec\{\bot\}$
\item $\mathcal{P}\conscrwl X\to X$ and it is clear that $\mathcal{P}\conscrwlf
  X\rec\{ X\}$
\item $\mathcal{P}\conscrwl c\to c$ with $c\in DC^0$ and it is clear that
  $\mathcal{P}\conscrwlf c\rec\{ c\}$
\end{itemize}

\noindent $\underline{l \Rightarrow l+1}$: the derivation can be of the following four forms:
\begin{itemize}

\item $\mathcal{P}\conscrwl \widehat{c(e_1,...,e_n)}\to c(t_1,...,t_n)$ by rule 3 of 
  {\it CRWL} and then we have $\mathcal{P}\conscrwl \hat{e_i}\to t_i$ for all $i\in\{
  1,...n\}$. By i.h. we have $\mathcal{P}\conscrwlf e_i\rec\cb_i$ with
  $t_i\in\cb_i$ and by rule 3 of  {\it CRWLF} we have $\mathcal{P}\conscrwlf
  c(e_1,...,e_n)\rec\mathcal{C}$ with $c(t_1,...,t_n)\in\mathcal{C}$.   

\item $\mathcal{P}\conscrwl \widehat{f(e_1,...,e_n)}\to t$, then there must exist 
  a rule $R=(f(\overline{s})\eqr e\Leftarrow \overline{C})\in\mathcal{P}$
  and $\theta\in CSubst_{\bot}$ such that by rule 4 of {\it CRWL} we will have the
  derivation

   $\mifrac{\hat{e_1}\to s_1\theta\quad ...\quad \hat{e_n}\to s_n\theta\quad e\theta\to
    t\quad \overline{C}\theta}
    {\widehat{f(e_1,...,e_n)}\to t}$
  By i.h. we have:
  \begin{itemize}
  \item[$i)$] there exists $\cb_i$ such that $\mathcal{P}\conscrwlf e_i\rec\cb_i$
    with $s_i\theta\in\cb_i$
  \item[$ii)$] there exists $\cb'$ such that $\mathcal{P}\conscrwlf
    e\theta\rec\cb'$ with $t\in\cb'$
  \item[$iii)$] $\mathcal{P}\conscrwlf\overline{C}\theta$
  \end{itemize}
  From $ii)$ and $iii)$, by rule 6 of {\it CRWLF} we can build the derivation
  $\mathcal{P}\conscrwlf f(s_i\theta)\recr\cb'$ using the c-instance
  $R\theta$. With this derivation and $i)$ we have $\mathcal{P}\conscrwlf
  f(\overline{e})\rec\cb$ such that $\cb'\subseteq\cb$, so $t\in\cb$.

\item $\mathcal{P}\conscrwl \hat{e}\con \hat{e}'$, using the rule 5 of {\it
    CRWL}. It follows that $\mathcal{P}\conscrwl \hat{e}\to t$ and
  $\mathcal{P}\conscrwl \hat{e}'\to t$ for some $t\in 
  CTerm$. By i.h. $\mathcal{P}\conscrwlf e\rec\cb$ and $\mathcal{P}\conscrwlf e'\rec\cb'$
  where $t\in\cb\cap\cb'$. Taking into account that $t\downarrow t$ for all
  $t\in CTerm$, by rule 9 of {\it CRWLF} we can build a derivation for 
  $\mathcal{P}\conscrwlf e\con e'$.

\item $\mathcal{P}\conscrwl \hat{e}\div \hat{e}'$, using the rule 6 of {\it
    CRWL}. It follows that $\mathcal{P}\conscrwl \hat{e}\to t$ and
  $\mathcal{P}\conscrwl \hat{e}'\to t'$ where $t,t'\in 
  CTerm_{\bot}$ and have a $DC$-clash. By
  i.h. $\mathcal{P}\conscrwlf e\rec\cb$ and $\mathcal{P}\conscrwlf e'\rec\cb'$ 
  where $t\in\cb$ and $t'\in\cb'$. By definition of $\uparrow$ and by rule 10 
 of {\it CRWLF} we can build a derivation for $\mathcal{P}\conscrwlf e\div e'$. 
$\mathproofbox$
\end{itemize}
\end{proof*}

\medskip

All the previous results make easy the
task of proving that we have done things right with respect to failure. 
We will need a result stronger than Prop. \ref{prop:4}, which does not provide
enough information about the relation between the denotation of an expression
and each of its calculable {\it SAS}'s.

\begin{proposition} 
\label{prop:5}
Given  $e\in Term_{\bot,\fail}$, if $\mathcal{P}\conscrwlf e\rec\cb$ and
$\mathcal{P}\conscrwl \hat{e}\to t$,
then  there exists $s\in\cb$ such that $s$ and $t$ are consistent.
\end{proposition}

\begin{proof}
  Assume $\mathcal{P}\conscrwlf e\rec\cb$ and $\mathcal{P}\conscrwl \hat{e}\to
  t$. By part $b)$ of Prop. \ref{prop:4} there exists $\cb'$  such that
  $\mathcal{P}\conscrwlf e\rec\cb'$ with $t\in\cb'$. 

  By Theorem \ref{th:1} it follows that $\cb$ and $\cb'$ are consistent. By
  definition of consistent {\it SAS}'s, as 
  $t\in\cb'$, then there exist $s\in\cb$ such that $t$ and $s$ are consistent.  
\end{proof}

\medskip

We easily arrive now at our final result.
 
\begin{theorem}
\label{th:2}
 Given  $e\in Term_{\bot,\fail}$,  if $\mathcal{P}\conscrwlf e\rec\{\fail\}$ then $[\! [ \hat{e}]\!
]=\{\bot\}$ \end{theorem}

\begin{proof}
  Assume $t\in [\! [ \hat{e}]\! ]$. This means that $\mathcal{P}\conscrwl
  \hat{e}\to t$, which in particular implies $t\in CTerm_{\bot}$. On the other hand, since
  $\mathcal{P}\conscrwlf e\rec\{\fail\}$, we know  from Prop. \ref{prop:5} that
  $\fail$ and $t$ must be consistent. As $\fail$ is consistent only with $\bot$
  and itself, and $t\in CTerm_{\bot}$, we conclude that  $t=\bot$.    
\end{proof}

\section{Final Discussion  and Future Work}
\label{sec:concl}

We have investigated how to deduce
negative information from a wide class of functional logic programs.
This is done by considering failure
 of reduction to head normal form, a notion playing a similar role, in the {\it FLP}  setting,
to that of negation as failure in logic programming, but having quite a different
starting point. Negation as failure in {\it LP}  can be seen mainly as an operational idea (existence
of a finite, failed search tree) of which a logical interpretation can be given
(successful negated atoms are logical consequences of the completion of the program).
The operational view of negation leads to an immediate implementation technique for negation
included in all Prolog systems:
to solve the negation of a goal, try to solve the goal and succeed if this attempt ends in failure.
Unfortunately, as it is well-known,
this implementation of negation is logically sound only for ground goals (see e.g. \citeN{AptBol94}).

Our approach has been different: we have given a logical status to failure
by proposing the proof calculus {\it CRWLF}
(Constructor based ReWriting Logic with Failure), which allows to deduce
failure of reduction within {\it CRWL}
\cite{GHL96,GHL99}, a well established theoretical framework for {\it FLP}.

We must emphasize the fact that {\it CRWLF} is {\em not} an operational mechanism
for executing programs using failure, but a deduction calculus fixing the logical meaning of
such programs. Exactly the same happens in \cite{GHL96,GHL99} with the proof calculus of {\it CRWL}, which
determines the logical meaning of a {\it FLP}  program, but not its execution.
The operational procedure in {\it CRWL} is given by a narrowing-based goal solving calculus,
which is proved to be sound and complete with respect to the proof calculus.
Our idea with {\it CRWLF} is to follow a similar way: with the proof calculus as a guide,
develop a narrowing-based operational calculus able to compute failures (even in presence of variables).
We are currently working on this issue.

It is nevertheless interesting to comment that the operational approach to failure mentioned
at the beginning of the section
for the case of Prolog, can be also adopted for {\it FLP}, leading to a very easy implementation of failure:
to evaluate ${\it fails}(e)$, try to compute a head normal form of $e$; if this fails, return {\it true},
otherwise return {\it false}. This is specially easy to be done in
systems having a Prolog-based implementation like {\it Curry} or \toy. We have checked that
all the examples in Section \ref{sec:interest} are executable in \toye with this implementation of failure,
if the function ${\it fails}$ is only applied to ground expressions. For instance, the goal
${\it safe}(c) \con T$ succeeds with answer $T=true$, and ${\it safe}(a) \con T$ succeeds with answer $T=false$.
If ${\it fails}$ is applied to expressions with variables, this implementation
is unsound. For instance, the goal ${\it safe}(X) \con false$ succeeds without binding $X$, which
is incorrect. The relationship between this kind of failure and {\it CRWLF}
is an interesting issue to
investigate, but it is out of the scope of this paper.

The most remarkable technical insight  in {\it CRWLF}
has been to replace the statements $e \to t$ of {\it CRWL} (representing a single reduction
of $e$ to an approximated value $t$)
 by $e \rec \cb$ (representing a whole, somehow complete, set $\cb$
of approximations to $e$). With the aid of $\rec$ we have been able
to cover all the derivations in {\it CRWL}, as well as to prove failure of
reduction and, as auxiliary notions, failure of joinability and divergence,
the two other kinds of statements that {\it CRWL} was able to prove.

The idea of collecting into an {\it SAS} values coming from
different reductions for a given expression $e$ presents
some similarities with abstract interpretation which, within the {\it FLP}  field,
has been used in \citeN{BertEchahed95} for detecting unsatisfiability of equations
$e = e'$ (something similar to failure of our $e \con e'$).
We can mention some differences between our work and \citeN{BertEchahed95}:

\begin{itemize}
\item[$\bullet$] Programs in \citeN{BertEchahed95} are much more
restrictive: they must be confluent, terminating, satisfy a property of
stratification on conditions, and  define strict and total functions. %\\[1ex]

\item[$\bullet$] In our setting, each {\it SAS} for an expression $e$ consists of
  (down) approximations to the denotation of $e$, and the set of {\it SAS}'s for $e$
  determines in a precise sense (Propositions \ref{prop:4} and \ref{prop:5}) the
  denotation of $e$. In the abstract interpretation approach one
  typically obtains, for an
  expression $e$, an abstract term representing a {\em superset} of the
  denotation of all the instances of $e$. But some of the rules of the
  {\it CRWLF}-calculus (like (9) or (10)) are not valid if we replace {\it SAS}'s by
  such supersets.
  To be more concrete, if we adopt an abstract interpretation view of our {\it
    SAS}'s, it  would be natural to see $\perp$ as standing for the set of all
  constructor terms (since $\perp$ is refinable to any value), and therefore to
  identify an {\it SAS} like ${\cal C} = \{\perp,z\}$ with ${\cal C}' = \{\perp\}$. But from   
  $e \rec {\cal C}$ we can deduce $e \con z$, while it is not correct to do the
  same from $e \rec {\cal C}'$.
  Therefore, the good properties of {\it CRWLF} with
  respect to {\it CRWL}
  are lost.

\end{itemize}

We see our work as a step in the research of a whole framework for
dealing with failure in {\it FLP}. Some natural future steps are
to develop model theoretic and operational semantics for programs making use
of failure information. On the practical side, we are currently working on an
implementation of failure for the {\it FLP} system \toye \cite{LS99a,Toy}.

\end{document}